\documentclass[twocolumn]{aastex701}

\usepackage{xcolor}
\usepackage{textgreek}
\usepackage[utf8]{inputenc}
\usepackage[english]{babel}

\usepackage{color,colortbl}
\definecolor{linkcolor}{rgb}{0.0,0.3,0.5}
\usepackage{tensind}
\tensordelimiter{?}
\DeclareGraphicsExtensions{.bmp,.png,.jpg,.pdf}
\usepackage{verbatim}
\usepackage[normalem]{ulem}
\usepackage{orcidlink}
\usepackage{soul}

\usepackage[T1]{fontenc}
\usepackage{amsmath,float} 
\usepackage{amssymb}

\begin{document}

\title{Tidal heating in detached double white dwarf binaries}

\author[orcid=0000-0003-1752-4882,sname='McNeill']{Lucy O. McNeill}
\affiliation{Department of Astronomy, Kyoto University, Kyoto 606-8502, Japan}
\affiliation{Hakubi Centre for Advanced Research, Kyoto University, Kyoto 606-8317, Japan}\
\affiliation{Center for Interdisciplinary Theoretical and Mathematical Sciences (iTHEMS), RIKEN, Saitama 351-0198, Japan}
\email[show]{mcneill@kusastro.kyoto-u.ac.jp} 

\author[orcid=0000-0002-8032-8174,sname='Hirai']{Ryosuke Hirai}
\affiliation{Astrophysical Big Bang Laboratory (ABBL), Pioneering Research Institute (PRI), RIKEN, Saitama 351-0198, Japan}
\affiliation{School of Physics and Astronomy, Monash University, Clayton, Victoria 3800, Australia}
\affiliation{OzGrav: The ARC Centre of Excellence for Gravitational Wave Discovery, Australia}
\email[]{ryosuke.hirai@monash.edu} 

\begin{abstract}
Short--period ($P<$1 hr orbits) detached double white dwarf binary (DWDB) components identified with transient surveys (e.g. SDSS, ZTF) have hot surface temperatures ($>$10,000~K) and observed radii a factor two larger than completely degenerate white dwarfs. We formulate tidal heating in helium composition extremely low mass white dwarf (ELM WD) components of detached DWDBs which reach mass transfer within a Hubble time. We {combine} a mass radius relation which varies with surface temperature and the {equilibrium} tidal friction model of Hut 1981, where the additional orbital energy loss from tidal friction is accounted for by increases in the primary surface temperature, and hence increasing radius. Applying this heating model to the current sample of binaries with ZTF, we predict temperature increases from the present day of up to $\sim$40\% before the onset of mass transfer. We find that helium white dwarfs are generically hot and large at the onset of mass transfer, even for the oldest DWDBs whose components can cool to be degenerate by the present day. In the population of Galactic DWDBs, we find that the onset of mass transfer should occur at orbital periods as long as 1000s (17 minutes), or binary gravitational wave frequency of 2~mHz. This is over three times longer than periods expected for degenerate WD (5 minutes). Since mass transferring DWDBs are progenitors for a variety of transients and stellar populations e.g. RCrB stars, AM CVn binaries, so-called type .Ia supernova, the finite temperature of donor white dwarfs should be taken into account. 
\end{abstract}


\section{INTRODUCTION}
Degeneracy pressure supported white dwarfs (WDs) are the expected remnants of the majority of stars, with millions to billions of double white dwarf binaries (DWDBs) thought to exist in the Milky Way \citep{Han1998,Nelemans2001a}. The DWDBs which form with short orbital periods $\lesssim$ hours, called ``short period DWDBs'', will reach {Roche lobe overflow} within a Hubble time due to the emission of gravitational waves. Short period DWDB formation requires mass exchange processes which can efficiently shrink the orbital separation before double compact object formation, namely common envelope (CE) ejection and stable mass transfer (MT).
\\

The majority of short period DWDBs contain at least one degenerate Helium white dwarf component \citep[e.g.,][]{Iben1997}. These are typically referred to as ``extremely low mass'' or ``ELM'' white dwarfs \citep{Brown2010}, with masses below around $0.3M_\odot$. In a DWDB, once the larger ELM WD component reaches Roche lobe overflow and begins donating mass, whether or not the binary merges depends \citep[at least in part;][]{Shen2015} on the stability of mass transfer \citep{Soberman1997}. Diversity in the component masses, composition, and final fates of the binary naturally leads to a variety of exotica. This includes hydrogen deficient AM Canum Venaticorum (AM CVn) binary systems, hydrogen poor stars
including R Coronae Borealis (RCrB) stars and subdwarfs (sdO, sdB) \citep{Webbink1984}, and also transients such as faint Type
.Ia supernovae \citep{Bildsten2007}, and calcium-rich {gap} transients \citep{Perets2010}. 
\\

Observing detached DWDBs offers a way to place constraints on the rates of the final outcomes, e.g. DWDB merger rates. A population of nearby detached short period DWDBs has been discovered by the ELM survey \citep{Brown2010}, and more recently with the Zwicky Transient Facility (ZTF). Of the known DWDB population with an ELM WD component, a small fraction are detected via eclipsing methods \cite[e.g.][]{Brown2011,Burdge2019a,Burdge2019b,Burdge2020,Burdge2020b,Kosakowski2023}. Eclipsing is not only advantageous in constraining the properties of the companion, but also in precisely measuring the radii of both components. This eclipsing population reveals hot, large primary ELM WD components\footnote[1]{Defined as the larger, in most cases less massive component.}. These primaries are typically at least twice {as large as that of fully degenerate WDs.} 
\\

This suggests that these ELM WD components are inflated by thermal effects and are only partially degenerate. One reason could be the internal dissipation caused by tidal interactions. For short period DWDBs, the orbital decay of the binary is dominated by energy and angular momentum loss from the emission of gravitational waves. However, it is suggested empirically that tidal heating has a $\sim10~\%$ contribution to the orbital decay \citep[e.g., in the binary ZTF J153932.16+502738.8,][]{Burdge2019a}. The perturbing tidal force depends sensitively on the ratio $R/a$ to the fifth power, and the radii $R$ of primaries in eclipsing ELM WD binary systems are $\gtrsim 10\%$ of their binary orbital separation $a$, so it is unsurprising that several binaries exhibit ellipsoidal variations in their light curves \citep{Hermes2014}. These variations are due to the tidally distorted shape of the larger primary{, substantiating the importance of tides in these systems}.
\\

The dissipation of orbital energy by tides in a binary tend to drive the component rotation frequencies to be synchronized with the orbital frequency, towards a circular orbit. 
The equilibrium tide model for tidal dissipation \citep{Hut1981,Eggleton1998} provides a formulation for the coupled response of the tidally distorted primary's quadrupolar hydrostatic shape/spin and the orbital separation (and eccentricity), due to dissipation by internal friction. This model has had success in a variety of astrophysics contexts, such as offering an explanation for so-called ``hot Jupiters'' and {the bimodal period distribution of Jupiter-mass planets} \citep[e.g.][]{Dawson2018}. 
\\

For ELM white dwarfs in DWDBs, \cite{Piro2011} considered tidal interactions in the 12.75 minute DWDB binary SDSS J065133.338+284423.37 \citep[hereafter J0651,][]{Brown2011}. J0651's orbital decay is primarily due to gravitational wave emission, but \cite{Piro2011} estimated that the contribution to the total orbital decay from tidal heating is $\sim5\%$. To do this, they parameterised the coupled orbital and spin evolution of the raised tidal bulge of the ELM WD component assuming that the orbital decay and angular momentum transfer from tides is facilitated by internal friction \citep{Goldreich1966}. This is ultimately very similar to the equilibrium tide model in that it is not necessary to model the detailed stellar structure to obtain the orbital decay rate from tidal heating. However, it does not account for the response of the ELM WD component to internal friction.
\\

{When there is tidal heating, there should be a non negligible increase in the surface luminosity.
This is because the heat deposited into the ELM WD via dissipative tidal interactions with the companion is comparable to ELM WD luminosities, and ELM WD radii are sensitive to the surface temperature. 
Work by \cite{DallOsso2014} modelled the ELM primary response to tidal heating in J0651 by modelling perturbations to polytropic white dwarfs with internal viscosity. 
They predicted that the radius responses by increasing up to around $\sim 5\%$ between now and Roche contact, and corresponding temperature increase up to $\sim 10\%$.}
\\

Another source of tidal heating is the damping of gravity modes (g--modes) in ELM white dwarfs. In a dynamical tide model, modes are excited in convectively stable regions by the orbit. Unlike the equilibrium tide model, which assumes a static bulge, the amplitude of the excited stellar oscillations (tides) depend very sensitively on the stellar structure, and also the orbital frequency. Specifically, strong g--mode excitation requires resonances. It is generally difficult for these low frequency dynamical oscillations to impact the orbital evolution. However, for DWDBs with ELM WD primaries, these frequencies are typically mHz, so that the energy dissipated can be comparable to or exceed predictions from equilibrium tide modes {\citep{Fuller2011,Burkart2013}} due to the frequency dependence of the excitation. With detailed stellar models \citep{Fuller2013}, recent work by \cite{Scherbak2024} includes the radius and temperature response to heating to model the future temperature evolution of the ELM WD component in in J0651. There, the g--mode excitation and heating necessarily requires detailed stellar evolution models. 
\\

In this paper, we formulate the increase in an ELM WD's luminosity from tidal heating {using an equilibrium tide model, which unlike g--mode models does not depend strongly on} detailed stellar structure. This generalization of the ELM WD's response to heating requires a temperature dependent mass radius relation for helium white dwarfs. We extend the equilibrium tide model of \cite{Hut1981} for the coupled orbital frequency and spin evolution, to predict temperature evolution of any primary ELM WD component in a short period DWDB. Our model can be readily applied to DWDBs with various primary masses and temperatures, because unlike dynamical tide models of g--modes it does not depend on orbital resonances with the mode frequencies, which depend sensitively on internal structure.
\\

 In Section~\ref{sec:sec2}, we present a mass radius relation for Helium WD based on existing numerical simulations \citep{Panei2000}. We also estimate the expected number of eclipsing binaries detectable by ZTF in Section~\ref{sec:sec3}. There we suggest that the hot and large ELM WD components in the eclipsing DWDB sample may be typical, motivating the following general tidal heating model. 
 In Section~\ref{sec:THmodel} we present the coupled temperature, frequency and primary rotation evolution for ELM WD components in circular DWDBs. This includes the temperature dependent mass radius relation. In Section~\ref{sec:future-evolution}, we apply our model to the ELM WD components in the ZTF sample of binaries, predicting their future surface temperature evolution with frequency until Roche contact. Then we dedicate Section~\ref{sec:J1539} to considering the past evolution of binary J1539 \citep{Burdge2019a}, to comment on the DWDB formation scenarios allowed by our tidal heating model. In Section~\ref{sec:discussion}, we compare this work with the literature in more detail, and finally discuss implications for future electromagnetic and gravitational wave detections. We summarise our findings in Section~\ref{sec:sec6}.

\section{Detached double white dwarf binaries}
\label{sec:sec2}
\begin{figure}
   \centering
\includegraphics[width=0.48\textwidth]{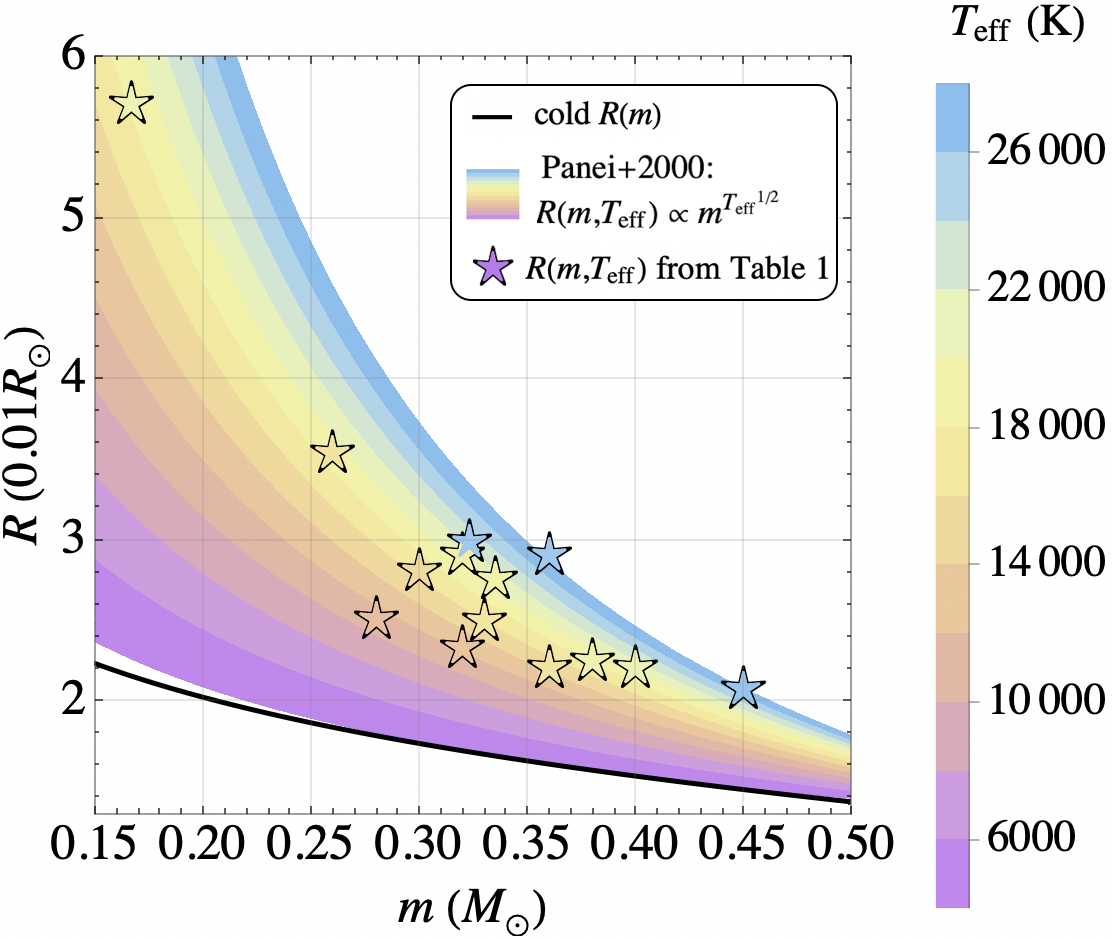}
   \caption{The masses and radii of eclipsing ELM WD binary components in the Galactic detached sample (Table~\ref{tab:WD1}) are shown by the stars, coloured according to their surface temperature. The majority of these ($m \lesssim 0.3 M_\odot$) are helium composition ELM WD. The secondary of J1539 which is likely CO composition is excluded, so there are a total of 14 stars in the plotting window. Comparing these stars to the completely degenerate (cold) relation (solid black line), the observed radii are typically larger by a factor between 2 to 3 for masses between 0.15--0.25$M_\odot$. Our empirical formulation of a temperature dependent mass radius relation are shown by the pastel contours, increasing in increments of 4,000~K. This relation is based on the helium composition models {which include a hydrogen envelope of mass $3 \times 10^{-4}M_\odot$} in \cite{Panei2000}, and is given by Equation~(\ref{eq:RMTPanei}).}
 \label{fig:mass-radius-temp}
 \end{figure}

The components of eclipsing binaries have inflated radii due to their hot surface temperatures.
In Table~\ref{tab:WD1}, we list select white dwarf components in eclipsing DWDBs discovered in SDSS and ZTF data \citep{Brown2011,Burdge2019a,Burdge2019b,Burdge2020,Burdge2020b}. 
\begin{table*}
\caption{Components from known DWDBs which have their mass $m$, radius $R$ and temperature $T_\mathrm{eff}$ constrained. They all have $T_\mathrm{eff}$ larger than 10,000~K, and are shown in Figure~\ref{fig:mass-radius-temp}. ZTF J1539+50271 has a asterix (*) since it is not shown in the plot.
When there are two components from the same binary, we label the larger component as the primary (rather than the hotter component, which is sometimes the convention). At present, there are a total fifteen components (from nine binaries) which have all three of their primary mass, radius and temperature measured. }
\label{tab:WD1}
\begin{center}
\begin{tabular}{ c c c c c c }
\hline
\textbf{ Name} & \textbf{Component} & \textit{\textbf{m}}($M_\odot$) & \textit{\textbf{R}}($R_\odot$)& \textit{\textbf{T}}$_\mathrm{eff}$(kK) & Ref. \\
 \hline
  ZTF J1539+50271* & Secondary &  $0.610^{+0.017}_{-0.022}$ & $0.01562^{+0.00038}_{-0.00038}$ &$48.9^{+0.9}_{-0.9}$ & \cite{Burdge2019a} \\
\hline
  
  ZTF J0538+1953 & Primary & $0.32^{+0.03}_{-0.03}$ & $0.02319^{+0.00032}_{-0.00026}$ & $12.8^{+0.2}_{-0.2}$  & \cite{Burdge2020b}\\
  
  ZTF J0538+1953 & Secondary & $0.45^{+0.05}_{-0.05}$ & $0.02069^{+0.00028}_{-0.00034}$ & $26.45^{+0.725}_{-0.725}$  & `` " \\
  \hline
  PTF J0533+02092 & Primary & $0.167^{+0.030}_{-0.030}$  & $0.057^{+0.004}_{-0.004}$ & $20^{+0.8}_{-0.8}$  & \cite{Burdge2019b} \\
  \hline
  ZTF J2029+1534 & Primary & $0.32^{+0.04}_{-0.04}$ & $0.029^{+0.002}_{-0.003}$ & $18.25^{+0.25}_{-0.25}$  & \cite{Burdge2020b}\\
  
 ZTF J2029+1534  & Secondary & $0.30^{+0.04}_{-0.04}$ & $0.028^{+0.003}_{-0.003}$ & $15.3^{+0.3}_{-0.3}$  &  `` "\\
  \hline
 ZTF J0722--1839  & Primary & $0.33^{+0.03}_{-0.03}$ & $0.0249^{+0.0001}_{-0.0003}$ & $16.8^{+0.15}_{-0.15}$  & `` " \\
  
  ZTF J0722--1839 & Secondary & $0.38^{+0.04}_{-0.04}$ & $0.0224^{+0.0004}_{-0.0002}$ & $19.9^{+0.15}_{-0.15}$  & `` " \\
  \hline
  ZTF J1749+0924 & Primary & $0.28^{+0.05}_{-0.04}$ & $0.025^{+0.004}_{-0.004}$ & $12.0^{+0.6}_{-0.6}$  & `` " \\
  
  ZTF J1749+0924 & Secondary & $0.40^{+0.07}_{-0.05}$ & $0.022^{+0.003}_{-0.004}$ & $20.4^{+0.2}_{-0.2}$ & `` " \\
  \hline
 ZTF J1901+5309 & Primary & $0.36^{+0.04}_{-0.04}$ & $0.029^{+0.001}_{-0.002}$ & $26.0^{+0.2}_{-0.2}$  & `` " \\
  
 ZTF J1901+5309 & Secondary & $0.36^{+0.05}_{-0.05}$ & $0.022^{+0.003}_{-0.002}$ & $16.5^{+2.0}_{-2.0}$ & `` "\\
  \hline
ZTF J2243+5242 & Primary & $0.323^{+0.065}_{-0.047}$ & $0.0298^{+0.0013}_{-0.0012}$ & $26.3^{+1.7}_{-0.9}$  & \cite{Burdge2020} \\
  
ZTF J2243+5242 & Secondary & $0.335^{+0.052}_{-0.054}$ & $0.0275^{+0.0012}_{-0.0013}$ & $19.2^{+1.5}_{-0.9}$ & `` " \\
  \hline
 SDSS J0651+2844  & Primary & $0.26^{+0.04}_{-0.04}$ &  $0.0371^{+0.0012}_{0.0012}$ & $16.53^{+0.2}_{-0.2}$  &  \cite{Hermes2012} \\
 \hline
\end{tabular}
\end{center}
\end{table*}
We select components with all three of mass $m$, radius $R$, and surface temperature $T_\mathrm{eff}$ measurements. In total, there are 15 components from 9 binaries, and the majority are helium composition ELM WD components. The mass (in solar mass $M_\odot$) and radii (in $0.01 R_\odot$) of these components are shown by the star shaped markers in Figure~\ref{fig:mass-radius-temp}. 
 These are coloured according to the measured $T_\mathrm{eff}$. This Figure does not show the first row in Table~\ref{tab:WD1} corresponding to J1539's secondary CO component. Instead, we dedicate Section~\ref{sec:J1539} to discussing J1539, where we speculate that the binary may not be detached. For most of this paper we are interested in detached binaries, so proceed for now by excluding J1539.
 
\subsection{Helium composition white dwarfs with finite surface temperature}
 
 All components in Figure~\ref{fig:mass-radius-temp} have measured radii at least 1.5 times the fully degenerate mass radius relation given by \citep{Verbunt1988}
\begin{equation}
\begin{split}
    & R_\mathrm{cold}(m) = 0.0114 \left( \left(m/14.4 \right)^{-2/3}- \left(m/14.4 \right)^{2/3}\right)^{1/2} \times \\
    & \left(1+ 3.5 \left( m/ \left( 5.7 \times 10^{-3}\right)\right)\right)^{-2/3} + \left( \left( 5.7 \times 10^{-3}\right)/m \right)^{-2/3}.
    \label{eq:egg}
    \end{split}
\end{equation} 
Here, $R_\mathrm{cold}$ is in units of $R_\odot$, and $m$ is the white dwarf mass in units $0.1 M_\odot$. This is shown by the solid black line in Figure~\ref{fig:mass-radius-temp}. 
To generalise the temperature dependent mass radius relation $R(m,T_\mathrm{eff})$ for the stars in Figure~\ref{fig:mass-radius-temp}, we refer to \cite{Panei2000} who modeled the cooling evolution of white dwarfs with various core compositions, including helium. They presented finite temperature dependent mass--radius relations for helium core composition white dwarfs with surface temperatures of 4, 8, 12, 16 and 20 kilo Kelvin (kK). We find that these five mass--radius relations (solid lines in their Figure 3{, corresponding to He WD with H envelopes $3 \times 10^{-4}M_\odot$}) are well described by a power law $\mathrm{d \ ln}R(m,T_\mathrm{eff})/\mathrm{d \ ln}m \propto -T_\mathrm{eff}^{1/2}$. The best fit for their mass radius relation is given by
\begin{equation}
R(m,T_\mathrm{eff}) = 0.0132 \times 10^{-0.177\sqrt{T_\mathrm{eff}}} \times m^{0.148 - 0.941\sqrt{T_\mathrm{eff}}} R_\odot.
\label{eq:RMTPanei}
\end{equation}
Here, $T_\mathrm{eff}$ is in units of 10,000~K, and $m$ is in units of 0.1$M_\odot$.
This relation, shown by the pastel contours in Figure~\ref{fig:mass-radius-temp} is consistent with the observed helium component radii in eclipsing detached binaries listed in Table~\ref{tab:WD1}. 

\subsection{Detached double white dwarf binary evolution}
For the short period double white dwarf binaries (DWDB) listed in Table~\ref{tab:binaries}, 
\begin{table*}
\caption{Binary evolution-related properties for the nine eclipsing binaries which have at least one of their components in Table~\ref{tab:WD1}. 
$f$ is the gravitational wave frequency, and $\mathcal{M}$ is the binary chirp mass. 
Assuming that each binary merges, $\tau_\mathrm{GW}$ is the time until merger from Equation~(\ref{eq:tGW}).
The Roche frequency $f_\mathrm{RL}(m_1)$ is found using Equation~(\ref{eq:fRL}), taking the {observed} present day temperature and primary (larger component) radius $R_1$. According to later sections, where we model the future heating (and hence radius increase) between now and Roche contact, this Roche frequency is an upper bound.
This frequency is used to calculate the time from the present day until Roche contact
$\tau_\mathrm{RL}$ with Equation~(\ref{eq:tRL}). We list the ratio $\tau_\mathrm{RL}/\tau_\mathrm{RL}$ in brackets, noting that when $R_1/a\gtrsim 0.1$, $\tau_\mathrm{RL}$ can be much shorter than $\tau_\mathrm{GW}$.
 $d$ is the distance in kpc, $a$ is the semi-major axis (separation), $P$ is the orbital period in seconds, and is equal to $2/f$. J1901's distance measurements are taken from Gaia \citep{Coughlin2020}. J1749's spectroscopic distance from \citep{Burdge2020}. All other distances are taken from Gaia DR3 \citep{Kupfer2024}.
}
\label{tab:binaries}
\begin{center}
\begin{tabular}{c c c c c c c c c c}
\hline
{Binary} & $f$(mHz) & $\mathcal{M}$ ($M_\odot$) & $\tau_\mathrm{GW}$ (Myr) & $f_\mathrm{RL}$(mHz) &  $\tau_\mathrm{RL}$ in Myr (\% $\tau_\mathrm{GW}$)  & $d$ (kpc) & $R_1/a$ & $P$ (s)  \\
 \hline
  ZTF J1539* & 4.8  &  0.303 (0.21 + 0.61) & 0.23 & {5.1} & {0.030 (13)}  & 2.469 $\pm$ 1.253 & 0.279 & 414.79  \\

  ZTF J0538 & 2.3 & 0.329 (0.32 + 0.45) & 1.41 & 10 & 1.38 (98) &  0.999  $\pm$ 0.366 & 0.129 &  866.60 \\

  PTF J0533 & 1.6 & 0.275 (0.167 + 0.652) &  4.89 &  {1.8} & {1.43 (29)} &  1.173 $\pm$ 0.390 & 0.245 & 1234.0 
  \\
  
  ZTF J2029 &  1.6 & 0.270 (0.32 + 0.3) & 5.26 & {7.6} & {5.13 (98)} &  1.095 $\pm$ 0.644 & 0.136  &  1252.1 \\
  
 ZTF J0722  & 1.4 & 0.310 (0.33 + 0.38) & 5.92  & {9.5} & {5.83 (99)} &  1.461 $\pm$ 0.785  & 0.102 & 1422.5\\
  
  ZTF J1749 & 1.3 & 0.290 (0.28 + 0.4) & 8.73 & {8.5} & {8.60 (99)} & $1.55^{+0.20}_{-0.18}$ & 0.0969 &   1586\\
  
 ZTF J1901 & 0.82 & 0.313 (0.36 + 0.36) & 24.1  & {8.0}  & {23.9} (100) & $0.915^{+0.96}_{-0.80}$ & 0.0833 & 2436  \\
  
ZTF J2243 & 3.8 & 0.286 (0.323 + 0.335) & 0.476 & {7.2}  & {0.39 (82)}  & 1.756 $\pm$ 0.726 & 0.244 &  527.93 \\
  
 SDSS J0651  & 2.6 & 0.311 (0.5+0.26) & 1.12 & {4.5} & {0.84 (76)} & 0.958  $\pm$ 0.370   & 0.215  &  765.21\\
 \hline
\end{tabular}
\end{center}
\end{table*}
energy and angular momentum loss from the emission of gravitational waves increases the gravitational wave frequency $f$ (related to orbital period via $P = 2/f$) according to \citep{Peters1964}:
\begin{equation}
 \left( \frac{\mathrm{d}f}{\mathrm{d}t}\right)_\mathrm{GW}  = 0.187 \left(\frac{\mathcal{M}}{M_\odot}\right)^{5/3} \left( \frac{f}{\mathrm{mHz}} \right)^{11/3} \mathrm{mHz / Myr}.
 \label{eq:dfdtGW}
\end{equation}
Here, $\mathcal{M}= \left( m_1 m_2 \right)^{3/5} \left( m_1 + m_2 \right)^{-1/5}$ is the chirp mass. Prior to mass transfer, the orbital decay of a detached binary proceeds approximately according to Equation~(\ref{eq:dfdtGW}) until Roche contact.  
In the absence of tides, this {can be approximated as} the time taken to get to infinite frequency from the present day $f$,
\begin{equation}
\tau_\mathrm{GW} = 2.00 \left( \frac{f}{\mathrm{mHz}} \right)^{-8/3} \left( \frac{\mathcal{M}}{M_\odot}\right)^{-5/3} \ \mathrm{Myr}.
\label{eq:tGW}
\end{equation}
Given the present day frequencies listed in Table \ref{tab:binaries}, Equation~(\ref{eq:dfdtGW}) drives the ELM WD binaries to mass transfer within kyr--Myr.
\\

 The onset of mass transfer occurs when the orbital separation $a$ is equal to the critical Roche Lobe radius $a_\mathrm{RL}$ \citep{Eggleton1983} for the larger (less massive) primary with mass $m_1$, which for ELM WD binaries is typically $\sim 3$ times the primary radius $R_1$. Written in terms of the binary mass ratio $q= m_1/m_2$, this is once the orbital separation $a$ has decreased to {$a_\mathrm{RL}$, which we take from \citep{Eggleton1983} 
  \begin{equation}
a_\mathrm{RL}/R_\mathrm{1} =   \frac{0.6q^{2/3}+\mathrm{log} \left(1+q^{1/3} \right)}{0.49 q^{2/3}} .
\label{eq:RRL}
 \end{equation} 
}
This is equivalently when the binary's gravitational wave frequency $f$ exceeds the Roche Lobe frequency $f_\mathrm{RL}$, given by {
\begin{equation}
    f_\mathrm{RL}(m_1) = \frac{G^{1/2} \left( m_1 + m_2 \right)^{1/2}}{\pi a_\mathrm{RL}^{3/2}}. 
    \label{eq:fRL}
\end{equation} 
In the approximation of $a_\mathrm{RL}$ by \cite{Pacynski1971}, $f_\mathrm{RL}$ is conveniently only a function of the primary $ \approx {2^{3/2}}/{9 \pi} \sqrt{{G m_1}/{R_1^3}}$ for mass ratios $q>0.52$.
} 
For each binary in Table~\ref{tab:binaries}, we take the larger ELM WD in Table~\ref{tab:WD1} and its observed $R$ as $R_1$ to calculate $f_\mathrm{RL}$ using Equation~(\ref{eq:fRL}). These are listed in Table~\ref{tab:binaries}. This finite frequency can be used to calculate the actual time, $\tau_\mathrm{RL}$ (shorter than $\tau_\mathrm{GW}$), taken from the present day frequency $f$ until mass transfer. Using Equation~(\ref{eq:dfdtGW}) this is given by
\begin{equation}
\begin{split}
    \tau_\mathrm{RL}&= \int_{f_\mathrm{RL}}^{f} \left( \frac{\mathrm{d}f}{\mathrm{d}t}\right)_\mathrm{GW}^{-1} \ \mathrm{d}f \\
    & = 2.00\left(\frac{\mathcal{M}}{M_\odot}\right)^{-5/3}\left( \frac{f_\mathrm{RL}^{8/3} -f^{8/3}}{f_\mathrm{RL}^{8/3} f^{8/3}}\right).
    \end{split}
    \label{eq:tRL}
\end{equation}
The time until mass transfer $\tau_\mathrm{RL}$ is listed in Table~\ref{tab:binaries}. The {ratio} of $\tau_\mathrm{RL}$ (calculated with Equation~(\ref{eq:tRL})) to $\tau_\mathrm{GW}$ from Equation~(\ref{eq:tGW}) is listed in parentheses in the same column. The majority of binaries have $\tau_\mathrm{RL}$ over 98\% $\tau_\mathrm{GW}$. In these cases, $\tau_\mathrm{RL}$ is roughly given by the merger timescale $\tau_\mathrm{GW}$, which is $\sim$Myr in these cases. However, for the four binaries where $R_1/a>0.2$ (J1539, J0533, J2243, J0651), $\tau_\mathrm{RL}$ is shorter by several 10's of percent. In particular, J0533 and J1539 are physically very close-- the binary separation to Roche radius is $a_\mathrm{RL}/a\gtrsim95\%$. The approximation that Roche contact time $ \tau_\mathrm{RL} \approx \tau_\mathrm{GW}$ is therefore valid as long as $f_\mathrm{RL}$ is more than a few times larger than the present-day frequency $f$, or as long as $R_1 \lesssim 0.1 a$. 

\subsection{Roche Lobe frequency marking the onset of mass transfer in ELM WD binaries}
\label{sec:RLfreq}
The times until Roche contact from the present day, $\tau_\mathrm{RL}$ listed in Table~\ref{tab:binaries} are all orders of magnitude shorter than the several Gyr long cooling timescales for ELM white dwarfs \citep{Mestel1952,Istrate2016}. Therefore for these short period binaries, their components of shown in Figure~\ref{fig:mass-radius-temp} will remain large and hot until the onset of mass transfer. If this luminous observed sample of ELM WDs in short period DWDBs is representative of the intrinsic short-period DWDB population, then the components in ELM WD binaries should be generically hot and large at the Roche contact frequency given by Equation~(\ref{eq:fRL}). This is in conflict with previous works \citep{Nelemans2001a} which assume that DWDBs are degenerate at Roche contact, due to being very old \citep{Tutukov1996}. For example, consider the lightest (and largest) ELM WD component $0.167M_\odot$ from J0533. Then, assuming the degenerate relation for the radius (Equation~(\ref{eq:egg})), the frequency that marks the beginning of mass transfer $f_\mathrm{RL}$ is 8.1~mHz by Equation~(\ref{eq:fRL}). However, using its actual inflated radius in Equation~(\ref{eq:fRL}) instead, it decreases to over four times smaller, 1.9~mHz. For most ELM WD in the sample, the radii are twice as large as the cold relation. In this case, the Roche frequencies $f_\mathrm{RL}$ are typically lower than the degenerate relation by a factor $2^{3/2} \approx 3$. {This stark decrease in the Roche contact frequency is expected even assuming that the radii of ELM WD do not increase between present day and Roche contact. We explore how possible radius expansion mechanisms could further influence the Roche contact frequencies in Section~\ref{sec:THmodel}. In any case, t}his may have consequences for mass transfer stability and related merger or transient rates for Galactic DWDBs. However, whether or not this observed population is representative of an intrinsic population is not clear. So we now estimate the total number of binaries expected from current eclipsing surveys.
\\

\subsection{Expected number of eclipsing DWDBs}
\label{sec:sec3}
Here we quantify whether the detected eclipsing binaries in Table~\ref{tab:binaries}, whose hot and large components are shown in Figure~\ref{fig:mass-radius-temp}, are likely typical or exceptional. The intrinsic number distribution $N(f)$ for a Galactic population of detached binaries, decaying toward mergers within a Hubble time via
Equation~(\ref{eq:dfdtGW}) is given by \citep[e.g.][]{Kyutoku2016}
\begin{equation}
\frac{\mathrm{d}N}{\mathrm{d}f} = \mathcal{R}_\mathrm{ELM}\left( \frac{\mathrm{d}f}{\mathrm{d}t}\right)_\mathrm{GW}^{-1}.
\label{eq:flux}
\end{equation}

Here, $\mathcal{R}_\mathrm{ELM}$ is the production rate of short period white dwarf binaries in the Milky Way. This rate is related to the star formation rate, which changes over Gyr timescales. The short period ELM WD binary population has gravitational wave frequencies $f\sim0.1$ to a few mHz, corresponding to Roche contact timescales $\tau_\mathrm{RL}\lesssim 10$Myr. Therefore the production rate $\mathcal{R}_\mathrm{ELM}$ can be taken as a constant in the mHz band. This is referred to as the ``steady-state'' assumption \citep{Kyutoku2016}, which for this rough calculation, we assume is valid over the whole range of frequencies listed in Table~\ref{tab:binaries}. This is also the frequency range over which mHz gravitational wave detectors such as LISA \citep{LISA2017} are most sensitive. 
\\

For a population of merging binaries with constant formation rate $\mathcal{R}_\mathrm{ELM}$, the number at a given frequency depends on the orbital decay rate for point masses $\left(\mathrm{d}f/\mathrm{d}t \right)_\mathrm{GW}$. However, this is strictly valid only for completely detached binary populations over the frequency range of interest. Binaries containing neutron stars and stellar mass black holes have $\sim 10$ km radii and merge in the $f\sim$~kHz range. They can be safely considered to be detached in the $f\sim$~mHz band, so that $\left(\mathrm{d}f/\mathrm{d}t \right)_\mathrm{GW}$ sets the whole population's number distribution in frequency space. Then Equation~(\ref{eq:flux}) can be used to infer the formation periods of Galactic binary neutron stars with the population detected by LISA \citep{Mcneill2022} if there is indication that $\mathcal{R}_\mathrm{ELM}$ is not a constant. On the other hand, DWDBs are thousands of times larger, and are expected to reach Roche lobe overflow in the mHz band. Even though $\mathcal{R}_\mathrm{ELM}$ may be a constant in the mHz band, in Section~\ref{sec:sec2} we suggest that from around 2~mHz the shape of the frequency distribution $N(f)$ will deviate from $\mathrm{d}N/\mathrm{d}f \propto \left(\mathrm{d}f/\mathrm{d}t \right)_\mathrm{GW}^{-1}$ due to mass transfer and mergers \citep{Seto2022}. Equation~(\ref{eq:flux}) is only valid when the whole intrinsic population can be considered detached. Our estimates for DWDBs in the mHz band that follows should therefore be taken as an upper bound.
\\

The electromagnetic detection of the ELM WD DWDBs we are considering suffers from various location-dependent selection effects such as extinction. Therefore Equation~(\ref{eq:flux}) cannot be immediately used for estimating the number of DWDBs expected in ZTF with $f\sim$~mHz in the same fashion as it can be used for estimates of gravitational wave detection. Electromagnetic selection effects must also be accounted for first. In addition to extinction, DWDB detection via eclipsing and ellipsoidal variations require specific orientations of the binary. Also, the cadence of ZTF limits detection via eclipsing to sufficiently short periods less than around 1 hour. Once we can quantify these three electromagnetic selection effects, we can use Equation~(\ref{eq:flux}) to obtain the number we expect to detect with e.g. ZTF based on a steady state assumption in the mHz band. Then we can finally compare this number to the observed sample of 9 binaries. Integrating Equation~(\ref{eq:flux}), and considering the various selection effects (characterized for now as ``selection''), we can estimate the number of binaries given some intrinsic production rate $\mathcal{R}_\mathrm{ELM} $ as
\begin{equation}
\begin{split}
N(>f_\mathrm{min}) & = \mathrm{selection} \times   \\ & \mathcal{R}_\mathrm{ELM}\int_{f\mathrm{min}}^{f\mathrm{max}} \left( \frac{\mathrm{d}f}{\mathrm{d}t}\right)_\mathrm{GW}^{-1} \ \mathrm{d}f.
\end{split}
\label{eq:NMW0}
\end{equation}
Here, $f_\mathrm{min}$ and $f_\mathrm{max}$ are the minimum and maximum frequencies over which our assumptions about detached binary evolution through frequency space hold. Cadence in eclipse detection should be considered for $f_\mathrm{min}$.
\\

For an intrinsic population of Galactic ELM WD binaries produced at a rate $\mathcal{R}_\mathrm{ELM}$, there is a limited
volume $V_\mathrm{obs}$ out to which ELM WD binaries can currently be detected. This is due a combination of the limiting magnitude of ZTF itself, and also Galactic extinction.
$V_\mathrm{obs}$ is only a small fraction of the total volume of the Milky Way $V_\mathrm{MW}$, so that the ratio ${V_\mathrm{obs}}/{V_\mathrm{MW}}$ roughly represents the detectable fraction of the Milky Way by eclipsing with ZTF. Since detection within this volume requires a specific binary orientation (inclination $i$), we must also include the probability, which we define as some function of the inclination $p(i)$. With these two fractions, Equation~(\ref{eq:NMW0}) becomes
\begin{equation}
N(>f_\mathrm{min}) = p(i) \times \frac{V_\mathrm{obs}}{V_\mathrm{MW}} \times \mathcal{R}_\mathrm{ELM} \int_{f\mathrm{min}}^{f\mathrm{max}} \left( \frac{\mathrm{d}f}{\mathrm{d}t}\right)_\mathrm{GW}^{-1} \ \mathrm{d}f.
\label{eq:NMW}
\end{equation}

For the probability that a binary has the necessary orientation for an eclipse $p(i)$, we assume that all binary inclinations are equally probable. Then $p(i)$ is the eclipsing probability for a binary with separation $a$ whose components have radii $R_1$ and $R_2$. This is $\left(R_1+R_2 \right)/a$. Ignoring the secondary radius $R_2$, we use the mean of this from the primaries in Table~\ref{tab:binaries}, and use that $p(i) = \langle R_1/a \rangle=$ 0.2. 
\\

 To estimate the observable volume $V_\mathrm{obs}$, we assume a cylindrical volume around Earth out to which ZTF can detect eclipsing ELM WD binaries in Table~\ref{tab:binaries}. For the radius $r_\mathrm{obs}$, we use the distance $d$ to the farthest binary where the primary is visible. 
This is J2243 at 1.756 kpc. Then, for the height $h$, we assume the denser Galactic thin disc to extend the scale height, which is 1kpc. From this, we construct a cylindrical observable volume $V_\mathrm{obs}= \pi r_\mathrm{obs}^2h = 9.7 \ \mathrm{kpc}^3$. Then, for the total Galactic volume $V_\mathrm{MW}$ we take a cylinder with the same height $h$, and a diameter of 30 kpc. Then $V_\mathrm{MW} = 709 \ \mathrm{kpc}^3$, resulting in an observable volume as a fraction of the total Galaxy of $ {V_\mathrm{obs}}/{V_\mathrm{MW}} = 0.014$. 
This estimate for $V_\mathrm{obs}$ as a fraction should be a good approximation within a few kpc, given various uncertainties and the small sample size. In addition, {we are assuming that the population density is uniform in the Galaxy.} However, if $r_\mathrm{obs}$ and hence the sample size increase, detailed Galactic density profiles \citep[e.g.,][]{Nelemans2001} can be used for more appropriate calculations of $ {V_\mathrm{obs}}/{V_\mathrm{MW}}$.
\\

Next in Equation~(\ref{eq:NMW}) is the intrinsic detached ELM WD binary production rate $\mathcal{R}_\mathrm{ELM}$. For this, we require an appropriate choice of an observationally motivated Galactic formation rate for the short period ELM WD binaries in Table~\ref{tab:binaries}. For more compact binaries such as binary neutron stars, which come into contact in the kHz band, the production rate is simply equal to the merger rate. This rate is deduced from LVK  detections, and also gamma ray bursts (GRBs). On the other hand, whether or not an ELM WD binary merges after it reaches Roche lobe overflow at $f_\mathrm{RL}\sim$~mHz depends on the stability of the mass transfer \citep{Soberman1997}. Depending on the responses of the donor star radius and orbital separation to mass transfer, stable mass transfer where $f \to 0$; $\dot{f}<0$ is possible so that the binary never merges, instead forming e.g. AM CVn systems. The other option is that the binary merges due to unstable mass transfer over dynamical timescales. {Since they do not necessarily merge, the formation rate should not be equated to the single merger outcome. The DWDB merger rate is therefore a lower limit for the formation rate $\mathcal{R}_\mathrm{ELM}$.}
\\

ELM DWDBs are candidate progenitors for several stellar populations, from which the merger rate of DWDBs containing ELM WDs may be inferred. R Coronae Borealis (RCrB) stars are a class of pulsating yellow supergiants, and likely the products of an ELM WD + CO WD merger. For the ELM WD binary merger rate in the Milky Way, the Galactic RCrB formation rate could be an suitable proxy. The population of RCrB stars implies an RCrB formation rate of $2-3 \times 10^{-3}$~yr$^{-1}$
\citep{Zhang2014,Karakas2015}. This is equivalent to an estimate of the merger rate of DWDBs containing an ELM WD component based on the ELM survey by \cite{Brown2016} of 0.003~yr$^{-1}$. On the other hand, the stably mass transferring AM CVn birthrate was also calculated with the ELM survey by \cite{Kilic2016}, and is 0.00017~yr$^{-1}$. Comparison between these two rates suggests that the majority of ELM WD binaries merge (by an order of magnitude). However, population synthesis studies \citep[e.g.][]{Kremer2015} predict that most Galactic DWDBs containing at ELM WD component become stably mass transferring AM CVn type systems. Given that there is no direct observational or theoretical proxy for the detached binary production rate $\mathcal{R}_\mathrm{ELM}$, we use these two rates as a range for $\mathcal{R}_\mathrm{ELM}$ in Equation~(\ref{eq:NMW}). 
\\

Next, for the minimum observable binary frequency $f_\mathrm{min}$, we note that the ZTF search pipeline which detected the majority of the sample \citep{Burdge2020b} is limited to variability with periods $P<1$~hr. This limits discoveries to binaries with $f > 0.6$ mHz. Also, since detection requires either an eclipse or ellipsoidal variations, we should consider frequencies where the binary radius $\langle R_1/a \rangle \gtrsim 0.1$. For ELM WD binaries, according to Table~\ref{tab:binaries} this is roughly when $f>$1 mHz. For this calculation, we take $f_\mathrm{min}=$ 1.3 mHz from J1749. This is the smallest frequency for binaries in Table~\ref{tab:binaries} which meet both of these criteria. For a maximum observable frequency $f_\mathrm{max}$, the integral in Equation~(\ref{eq:NMW}) is much less sensitive to $f_\mathrm{max}$ than it is to the minimum $f_\mathrm{min}$. We take $f_\mathrm{max}$ equal to 3.8 mHz, corresponding to J2243. But we note that the frequency distribution given by $\left(\mathrm{d}f/\mathrm{d}t \right)$ likely decreases when $f\gtrsim 2$~mHz due to binaries with the lowest primary masses $0.15-0.2M_\odot$ in the population undergoing mass transfer (Equation~(\ref{eq:fRL}) and Figure~\ref{fig:mass-radius-temp}).
\\

Then, for the frequency derivative term $\left({\mathrm{d}f}/{\mathrm{d}t} \right)_\mathrm{GW} $ in Equation~(\ref{eq:flux}), we use the mean binary chirp mass of the sample presented in Table~\ref{tab:binaries}, ${\mathcal{M}} = 0.3 M_\odot$. We can now estimate a range for the number of ELM double white dwarf binaries expected to be detected with the eclipsing ZTF pipeline. Using Equation~(\ref{eq:NMW}), the total number of detached binaries we expect from eclipsing $N_\mathrm{ZTF}$ is
\begin{equation}
\begin{split}
    N_\mathrm{ZTF} &= 
    42 \times \frac{\mathcal{R}_\mathrm{ELM}}{0.001 \mathrm{yr}^{-1}} \left( \frac{\mathcal{M}}{ M_\odot}\right)^{-5/3} \times 
    \\ & \left(\left( \frac{f_\mathrm{min}}{\mathrm{mHz}}\right)^{-8/3}-\left( \frac{f_\mathrm{max}}{\mathrm{mHz}}\right)^{-8/3}\right)
    \\&
    =  3-59\mathrm{\ binaries},
    \label{eq:NZTF}
    \end{split}
\end{equation} 
corresponding to the range bound between the two {$\mathcal{R}_\mathrm{ELM}$} rates of $1.7 \times 10^{-4}$ {(AM CVn binaries)} and $0.003 \ \mathrm{yr}^{-1}$ {(detached DWDBs, RCrB stars). We show $N_\mathrm{ZTF}$ given these two rates in Figure~\ref{fig:NZTF} with the black left hand vertical axis. 
\begin{figure}
  \centering
\includegraphics[width=0.5\textwidth]{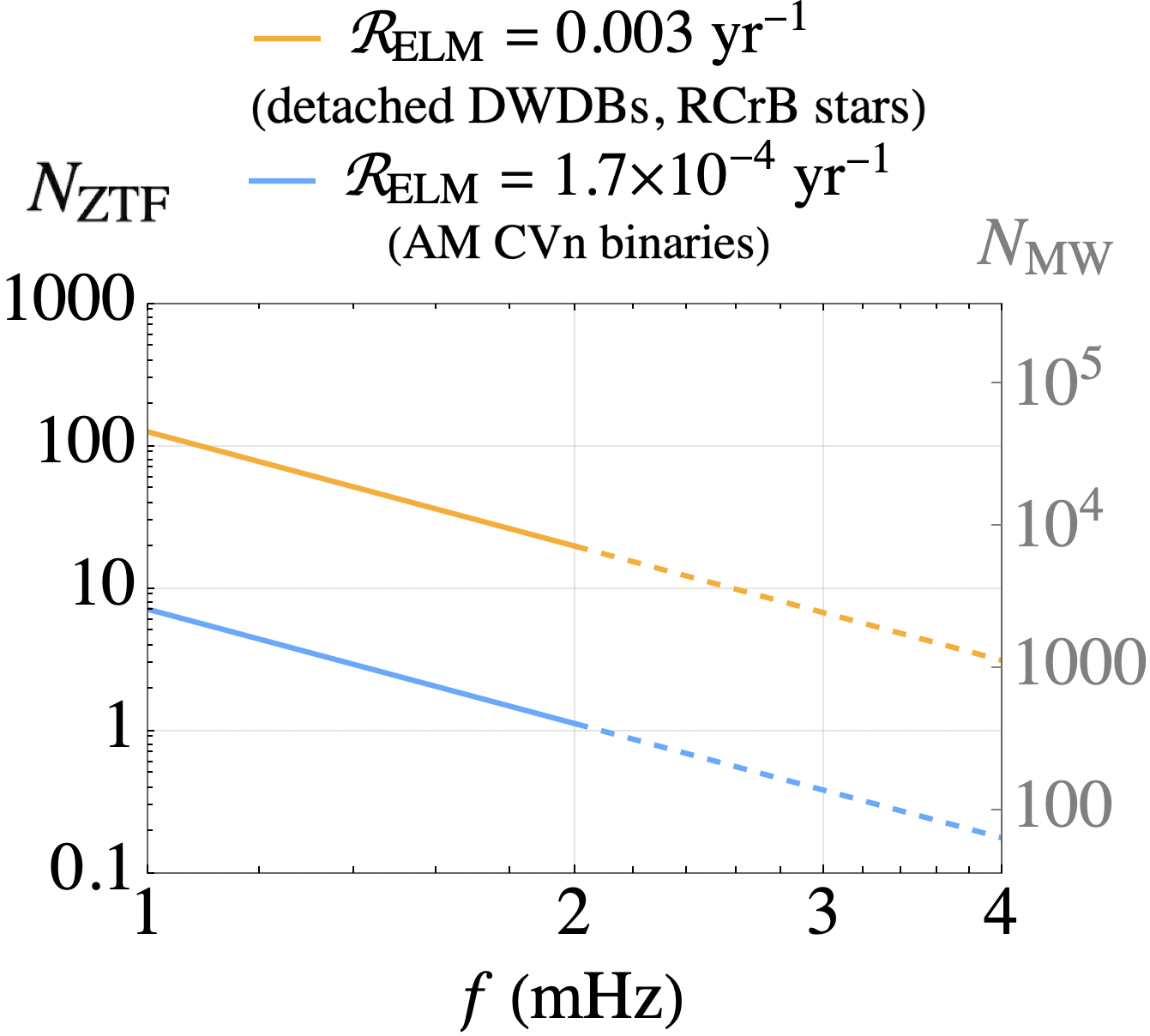}
  \caption{{Reverse cumulative distribution function for DWDBs as a function of $f$, using the estimate for ELM DWDB production rates $\mathcal{R}_\mathrm{ELM}$ via RCrB stars and also detached ELM DWDBs (orange) and AM CVn binaries (blue). The number expected by current surveys, $N_\mathrm{ZTF}$, which accounts for ELM selection effects via Equation~(\ref{eq:NZTF}) is given by the black vertical axis on the left hand side.
  On the right hand side axis in grey is the intrinsic population in the Galaxy, $N_\mathrm{MW}$ (Equation~(\ref{eq:NMW})). {These two axes differ by the product of the eclipsing probability and observable volume ${V}_\mathrm{obs}$.} Integrating from 1.3mHz to 3.8 mHz, the estimated $N_\mathrm{ZTF}$ are 3 (blue) and 59 (orange) respectively. The number in the observed sample (Table~\ref{tab:binaries}) is 9.}
  {For the dashed portions of the lines, we suggest that both $N_\mathrm{ZTF}$ and $N_\mathrm{MW}$ may no longer be proportional to $f^{-8/3}$ for $f>2$~mHz as discussed in Section~\ref{sec:population_discussion}.}
  }
\label{fig:NZTF}
\end{figure}
} 
\\

The {larger} estimate obtained with the merger rate from {RCrB} stars ({orange line}), 59 binaries, is several times larger than the number of binaries in the table (9 binaries). However, using the much smaller stably mass transferring AM CVn birthrate ({blue line}) for the binary production rate, the expected number is $N_\mathrm{ZTF}=3$ binaries. {The amplitude of $N_\mathrm{ZTF}$ (left vertical axis) in Figure~\ref{fig:NZTF} are set by $\mathrm{selection} \times \mathcal{R}_\mathrm{ELM}$, where $\mathrm{selection}$ depends} sensitively on the maximum distance ({volume }${V_\mathrm{obs}}\propto r_\mathrm{obs}^2$) {where we used the furthest distance in the sample. Then for the calculation of $N_\mathrm{ZTF}$, the integral depends sensitively on the chosen} minimum frequency ($ \propto f_\mathrm{min}^{-8/3}$){. Given our assumptions,} the sample size of nine (9) binaries falls within our predicted range {for $N_\mathrm{ZTF}$} from the two rates in Equation~(\ref{eq:NZTF}). 
\\ 

{We also show the corresponding intrinsic population $N_\mathrm{MW}$ which is computed with Equation~(\ref{eq:NMW}) with the right hand side vertical axis, marked by grey ticks. This estimate is made without EM selection effects and hence is relevant for DWDB detection by gravitational waves.} {We can compare $N_\mathrm{MW}$ for our two choices of $ \mathcal{R}_\mathrm{ELM}$ with estimates for He+He and He+CO DWDBs from simulations of the Milky Way \citep[e.g., Figure 7 in][]{Lamberts2019}. In their population synthesis estimates, $N_\mathrm{MW}$ follows the same power law $\propto f^{-8/3}$, but passes $\approx 10^4$ at $f=1$~mHz. Their curve therefore lies between our orange and blue curves. In addition to the observed ZTF sample, these two estimates also seem consistent with simulation-based estimates of the intrinsic population.}
According {to our} calculation, the observed population could be more or less complete, and are not strongly biased to brighter objects that have exceptionally high temperatures. {In this case, assuming that the temperature of ELM WD in DWDBs is only governed by cooling, this suggests typical ages of DWDBs $\lesssim$~Gyr.}
\\

{Finally we note the change in the line style from solid to dashed for frequencies $f>2$~mHz in Figure~\ref{fig:NZTF}. This is because we do not expect either the ZTF or intrinsic population's shape via $\mathrm{d}f/\mathrm{d}t$ to be given by Equation~(\ref{eq:dfdtGW}) when $f\gtrsim 2$~mHz.
There are two reasons for this, which are discussed in detail later in Section~\ref{sec:population_discussion}. One reason which we can note here is that the intrinsic population $N_\mathrm{MW}$ should drop off for frequencies where the onset of mass transfer in the population can occur. Table~\ref{tab:binaries} suggests that $f_\mathrm{RL}$ can be as low as 1.9~mHz. Therefore for frequencies $f\gtrsim 2$~mHz, in the whole population DWDBs can start disappearing in frequency space, leading to smaller $N_\mathrm{MW}$ there.
}

\section{Tidal heating model}
\label{sec:THmodel}
One possible explanation for the hot surface temperatures, and consequently large radii for the ELM WD components in Figure~\ref{fig:mass-radius-temp} and Table~\ref{tab:WD1} is tidal heating. In the equilibrium tide model, changes in primary rotation $\Omega_1$ and orbital separation $a$ occur over the tidal friction timescale $\tau_\mathrm{TF}$ (``slow'' timescale in \citealt{Goldreich1963,Hut1981}):
\begin{equation}
    \tau_\mathrm{TF} = \frac{1}{6} \frac{m_1}{m_2}  \frac{\mathcal{Q}}{k} \left( a/R_1\right)^5 P_\mathrm{force},
    \label{eq:tauTF}
\end{equation}
where the tidal quality factor $\mathcal{Q}$ \citep{Goldreich1966} is the number of dynamical oscillations after which a free oscillation would reduce its amplitude by half {via} internal friction. In other words, it is a dimensionless measure of the damping efficiency. The apsidal motion constant $k$ describes the coupling between the orbital and the excited mode in question. In an equilibrium tide model, it is natural to use the apsidal motion constant $k$ associated with the fundamental quadrupolar mode, which are most strongly coupled to the orbit. For degenerate white dwarfs, $k=0.14$ \citep{Sterne1939}. However, since finite temperature ELM WDs should be more centrally condensed than if they were completely degenerate, a more appropriate value is likely $k \lesssim 0.1$. The tidal friction (TF) timescale in Equation~(\ref{eq:tauTF}) {relates to} the instantaneous orbital decay rate from tidal heating
\begin{equation}
    \frac{1}{a} \left( \frac{\mathrm{d}a}{\mathrm{d}t} \right)_\mathrm{TH} =  \tau_\mathrm{TF}^{-1}.
    \label{eq:basic-timescale}
\end{equation}
The binary gravitational wave frequency $f \equiv\sqrt{G(m_1+m_2)/a^3}/(\pi)$, so that the frequency increase due to tidal heating is given by $ f^{-1} \left( \mathrm{d}f / \mathrm{d} t \right)_\mathrm{TH} = 3/2 \tau_\mathrm{TF}^{-1}$.
Given the strength of the tidal interaction and damping rate characterised by ${k}/{\mathcal{Q}}$, and taking the forcing period $P_\mathrm{force}$ to be the inverse of the mean motion $n=2 \pi f_\mathrm{orb} = \pi f$, the instantaneous increase in $f$ is given by
\begin{equation}
 \left( \frac{\mathrm{d}f}{\mathrm{d}t}\right)_{\mathrm{TH}} = \frac{9 k}{\mathcal{Q}} \frac{ m_2 \pi^{13/3}R_1^5 f^{16/3}}{G^{5/3} m_1(m_1+m_2)^{5/3}}.
 \label{eq:dfdtTD1}
\end{equation}
\\

For ELM white dwarfs, we can take $R_1$ from Equation~(\ref{eq:RMTPanei}) for any primary mass $m_1$ and surface temperature $T_\mathrm{eff}$. A suitable value for the ratio of $k/\mathcal{Q}$ can be inferred from \cite{Piro2011}, who considered {the current luminosity of the $0.26M_\odot$ ELM WD primary in J0651, and the limiting tidal heating efficiencies.} For a decaying binary, the increase in the timing of eclipses given an orbital period decrease $\dot{P}$ over a time interval of $t_\mathrm{obs}$ can be written in terms of the increase in gravitational wave frequency $\dot{f}$
\begin{equation}
\frac{\dot{P}}{P} \frac{t_\mathrm{obs}^2}{2} = - \frac{ \dot{f}}{f} \frac{t_\mathrm{obs}^2}{2}.
\end{equation}
 Considering only gravitational wave emission, the arrival time is predicted to increase by ${\dot{P}_\mathrm{GW}}/{P} \times t_\mathrm{obs}^2/2 = - \dot{f}_\mathrm{GW}/f \times t_\mathrm{obs}^2/2$ = 5.5 s earlier after $t_\mathrm{obs}=$ 1 year. However, after including an additional contribution from tidal heating, the eclipse occurs ${( \dot{P}_\mathrm{GW} + \dot{P}_\mathrm{TH} ) }/{P} \times t_\mathrm{obs}^2/2 = - ( \dot{f}_\mathrm{GW} + \dot{f}_\mathrm{TH})/f \times t_\mathrm{obs}^2/2$ = 5.8 s earlier. {This is in the limit that the primary is tidally locked, i.e. assuming relatively efficient tidal heating.} This predicted additional increase of 0.3 seconds means that $ \dot{f}_\mathrm{TH}$ in Equation~(\ref{eq:dfdtTD}) is 5.4\% of $\dot{f}_\mathrm{GW}$ in Equation~(\ref{eq:dfdtGW}). Solving this for $k/\mathcal{Q}$ with J0651's binary properties listed in Table~\ref{tab:binaries}, we find that $k/\mathcal{Q} =  8 \times 10^{-12}$. {In the other limit when it is not tidally locked, and the luminosity is primarily supplied by WD cooling not tidal interaction with its 0.5$M_\odot$ companion, then $\mathcal{Q}= 7 \times 10^{10}$. This value is an upper bound. {Taking} $k=0.1$, then we are assuming that $\mathcal{Q}= 1.3 \times 10^{10}$, which is a factor few times smaller than this upper bound.}
 We treat this {$k/\mathcal{Q}$} as universal across the ELM {WD} population, and proceed by calculating the tidal heating timescales in Equation~(\ref{eq:tauTF}). These are listed in Table~\ref{tab:timescales}.  \begin{table}
\caption{Tidal heating and cooling timescales for the primary components of the nine binaries. This is calculated using the same $k/\mathcal{Q}$ value that \cite{Piro2011} found for J0651. The ratio of $\tau_\mathrm{cool}/\tau_\mathrm{TH}$ is listed in the final column. This ratio is shown in logarithm base 10 for each binary in Figure~\ref{fig:timescale-ratio}.  For ratios much greater than 1, white dwarf cooling can be neglected. 
 All binaries (except for J1901) have $\tau_\mathrm{TH}< \tau_\mathrm{cool}$, so that their current temperature evolution is governed by tidal heating over Myr timescales{.}}
\label{tab:timescales}
\begin{center}
\begin{tabular}{ c c c c }
\hline
{Binary} & $\tau_\mathrm{TH}$ (Myr) & $\tau_\mathrm{cool}$ (Myr) & ratio  \\
 \hline
  ZTF J1539* & 4.71  &  5564.45 & 1180 \\

  ZTF J0538 & 733 & 9325.86 & 12.7 \\

  PTF J0533 & 11.3 & 3851.99  & 341
  \\
  
  ZTF J2029 &  1818 & 9325.86 & 5.13 \\
  
 ZTF J0722  & 4564 & 9410.13 & 2.06 \\
  
  ZTF J1749 & 5029 & 8830.64 & 1.76 \\
  
 ZTF J1901 & 36713 & 9652.56 & 0.26\\
  
ZTF J2243 & 18.6  & 9351.34 & 502 \\
  
 SDSS J0651  & 60.7 & 7840.16 & 129 \\
 \hline
\end{tabular}
\end{center}
\end{table}
For the ELM WD population found in eclipsing binaries, orbital evolution from tidal heating occurs over $\tau_\mathrm{TH}$ equal to a few Myr to several Gyr.

\subsection{White dwarf cooling vs tidal heating dominated temperature evolution}
\label{sec:ratios}
After formation, finite temperature white dwarfs cool as they radiate residual heat. This competes with any possible white dwarf heating process. For the cooling timescales for helium composition ELM white dwarfs, we use the modified Mestel cooling law in \cite{Hurley2003}, which gives the surface luminosity decrease as a function of time:
 \begin{equation}
 L = \begin{cases} 
      \frac{300 M_\mathrm{WD} Z^{0.4}}{\left(4\left(t+0.1 \right) \right)^{1.18}} & t\leq 9 \mathrm{\ Gyr} \\
      \frac{300(9000A)^{5.3}) M_\mathrm{WD} Z^{0.4}}{\left(4\left(t+0.1 \right) \right)^{6.48}} & t> 9 \mathrm{\ Gyr}.
   \end{cases}
   \label{eq:Lcool2}
\end{equation}
Here, the baryon number $A=4$ for Helium composition, and time $t$ is in Myr. Assuming that the white dwarf is a blackbody, in terms of the temperature and radius the luminosity can be expressed as $L=4 \pi \sigma R^2 T^4$, where $\sigma = 5.67 \times 10^{-5}~\mathrm{erg~cm^{-2}~K^{-4}~s^{-1}}$ is the Stefan-Boltzmann constant. For each of the binaries in Table~\ref{tab:binaries}, we take the radii and temperature of the primary components from Table~\ref{tab:WD1} for the initial luminosity. This is at $t=0$ in Equation~(\ref{eq:Lcool2}). Assuming that each ELM WD can be considered ``cold'' when its temperature drops to 4,000~K (see Figure~\ref{fig:mass-radius-temp}) , we solve for the time $t$ that it takes to cool to 4,000~K, and list these times in Table~\ref{tab:timescales}. These cooling timescales $\sim 4-10$~Gyr are consistent with detailed ELM WD cooling models \citep{Serenelli2001,Istrate2016}.
\\

\begin{figure}
  \centering
    \includegraphics[width=0.5\textwidth]{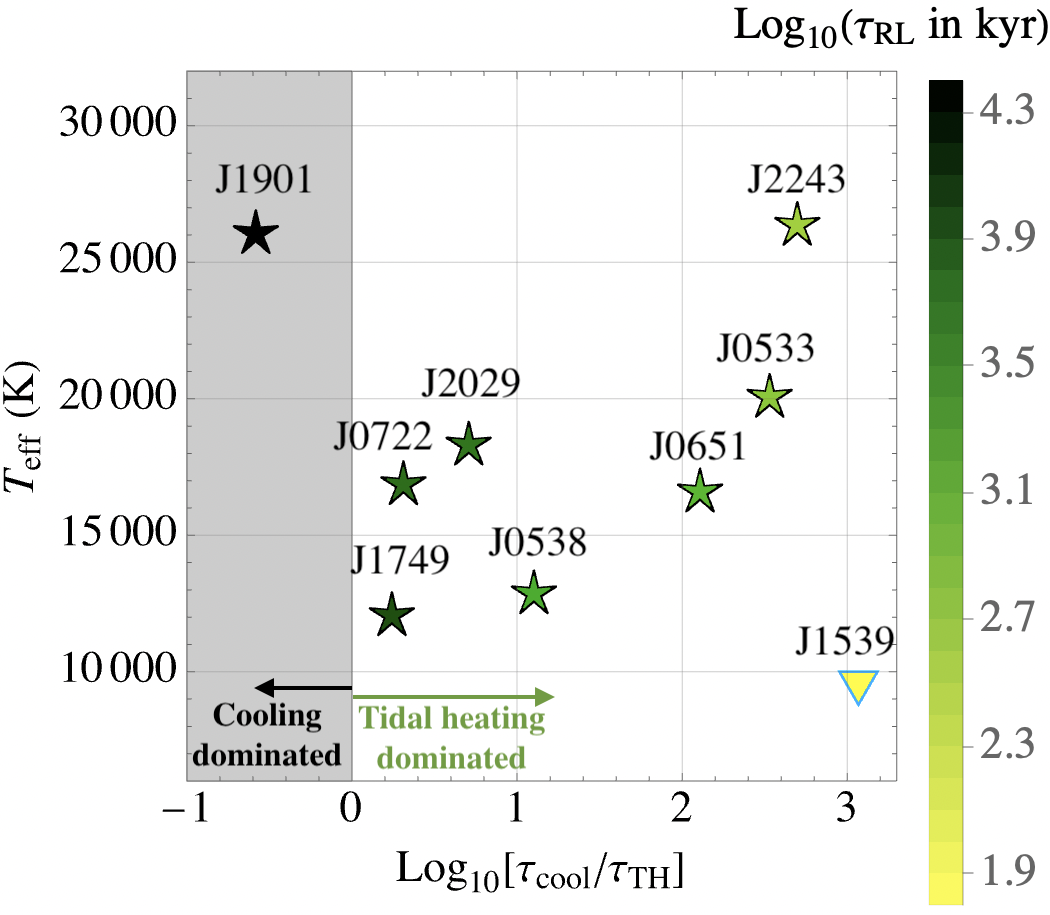}
  \caption{The ratio of the cooling timescale $\tau_\mathrm{cool}$ (via Equation~(\ref{eq:Lcool2})) to tidal heating timescale $\tau_\mathrm{TF}$ (Equation~(\ref{eq:tauTF})) given by $\tau_\mathrm{cool}/\tau_\mathrm{TH}$ is shown (log scale) against the surface temperature (linear scale). 
  At $\tau_\mathrm{cool}/\tau_\mathrm{TH}=1$ (Log$_{10}(\tau_\mathrm{cool}/\tau_\mathrm{TH} ) = 0$), we divide the horizontal axis into cooling and tidal heating dominated temperature evolution regimes. When $\tau_\mathrm{cool}/\tau_\mathrm{TH}$ is less than 1 (grey shaded area), the current temperature evolution {into the future} is governed by white dwarf cooling. 
  For each primary component in the binary population, its ratio from Table~\ref{tab:timescales} and temperatures (Table~\ref{tab:WD1}) are shown by the star shaped markers. J1539's primary (not in Table~\ref{tab:WD1}) is included as the blue arrow symbol with its temperature upper bound. 
Components are shaded according to their time until Roche contact $\tau_{\mathrm{RL}}$, driven primarily by the emission of gravitational waves, using a yellow to black colour scheme. 
}
\label{fig:timescale-ratio}
\end{figure}

With the two timescales associated with the competing heating and cooling processes listed in Table~\ref{tab:timescales}, we can now determine whether the ELM WD primary in each binary's temperature evolution is currently governed by cooling or tidal heating. We calculate the ratio of heating to cooling timescales $\tau_\mathrm{cool}/\tau_\mathrm{TH}$  (Equations~(\ref{eq:Lcool2}) and (\ref{eq:tauTF})) and list them in Table~\ref{tab:timescales}. We plot this ratio with star shaped markers on a logarithmic (base 10) scale in the horizontal axis of Figure~\ref{fig:timescale-ratio}.
The vertical axis is for the current primary temperatures on a linear scale. When $\tau_\mathrm{cool}/\tau_\mathrm{TH}<1$ (or $\mathrm{log}_{10} \left(\tau_\mathrm{cool}/\tau_\mathrm{TH}\right)<0$), the component's current surface temperature evolution $\dot{T}_1$ is cooling dominated via Equation~(\ref{eq:Lcool2}). This region is shaded in grey. When $\tau_\mathrm{cool}/\tau_\mathrm{TH}>1$ (unshaded area), the component's current (and future) $\dot{T}_1$ is tidal heating dominated. According to our model, eight out of nine binaries from Table~\ref{tab:WD1}, except for J1901 currently have temperatures governed by tidal heating. We also included the ratio for the $0.21M_\odot$ primary component in J1539 shown by the blue upper-bound arrow symbol. This low mass primary component of the binary was not included in Table~\ref{tab:WD1} since its temperature is not actually measured. But it has a reported upper bound of 10,000~K \citep{Burdge2019a}. Using its measured radius $0.0314 R_\odot$, J1539's primary has a ratio log$_{10}\left(\tau_\mathrm{cool}/\tau_\mathrm{TH}\right)=$ 3.1. This component has the highest ratio, and also the coolest temperature. {The primaries are shaded according to the log$_{10}$ of their time until Roche contact (in kyr), which are all $\sim$~Myr. Since this is $\ll \tau_\mathrm{cool}$, {we can safely} ignore WD cooling for all nine binaries} when we consider {their} future evolution later in {Sections~\ref{sec:future-evolution} and \ref{sec:J1539}}.
\\

In Figure~\ref{fig:timescale-ratio} the horizontal axis can equivalently be thought of as a measure of time. For a given binary, this ratio log$_{10}\left(\tau_\mathrm{cool}/\tau_\mathrm{TH}\right)$ will increase as the orbital separation shrinks primarily due to the emission of gravitational waves. Specifically, this increase in the ratio is due to the strong dependence on the separation via $\left( a/R_1\right)^5$ in Equation~(\ref{eq:tauTF}). Although the sample size is small, there appears to be a positive correlation between ratio and surface temperature. In other words, primaries which are at later times in their binary evolution, or primaries in binaries which are generally closer to Roche contact, seem to have larger temperatures. 
\\

In the tidal heating dominated temperature evolution regime (unshaded region), this positive correlation is expected anyway. In Equation~(\ref{eq:tauTF}), the separation $a$ decreases in time primarily due to the emission of gravitational waves. The smaller $a/R_1$ leads to larger energy dissipation rates inside the primary by tidal heating via Equation~(\ref{eq:dfdtTD1}). This internal friction should be accounted for in the primary surface luminosity, which means that the temperature $T_1$ should increase. In addition, $R_1$ increases as a response to the tidal heating, but these two effects become clearer when we formulate the temperature evolution later in Section~\ref{sec:future-evolution}. In short, both effects contribute to decreasing $a/R_1$ in $\tau_\mathrm{TH}$ as the binary evolves in time, accelerating the damping rate and hence temperature increase.
\\

{Detached binaries evolve in time via a decreasing $a/R_1$, from left to right in Figure~\ref{fig:timescale-ratio}. In the absence of tidal heating, we expect no correlation between temperature $T_1$ and time. 
This is because for all binaries, the time until Roche contact is $\lesssim10$~Myr, much shorter than the Gyr cooling timescales via Equation~(\ref{eq:Lcool2}). 
On the other hand, if tidal heating is possible as binaries evolve towards Roche contact, we expect to see a positive correlation between $T_1$ and the ratio log$_{10}\left(\tau_\mathrm{cool}/\tau_\mathrm{TH}\right)$ in Figure~\ref{fig:timescale-ratio}. Especially later in the evolution when log$_{10}\left(\tau_\mathrm{cool}/\tau_\mathrm{TH}\right) \gtrsim 100$, $\tau_\mathrm{TH}$ becomes comparable to the inspiral timescale $\tau_\mathrm{RL}$. There we expect a relatively sudden non-negligible temperature increase from tidal heating. Our tidal heating timescale therefore seems qualitatively consistent with this correlation in the observations.}
\\

{Although the majority of binaries are currently in the tidally heated regime, their present day temperatures are still primarily from residual heat. We stress that it is not possible to make the observed ELM WD temperatures by assuming e.g. DWDB formation 10~Gyr ago ($f\sim0.1$~mHz), where components can cool to 4,000~K by observed frequencies $f\sim1$~mHz to be tidally heated. Instead, the progenitor DWDBs can only cool to around 10,000~K, to possibly be tidally heated. Binaries with log$_{10}\left(\tau_\mathrm{cool}/\tau_\mathrm{TH}\right) \gtrsim 100$ may get noticeably hotter in the future before reaching Roche contact (see later in Section~\ref{sec:future-models}). We already suggested the possibility that the observed sample is a complete sample based on our estimates for $N_\mathrm{ZTF}$ at the end of Section~\ref{sec:sec2}. Given the results of this section, it might be common that DWDBs form with merger timescales $\lesssim1$~Gyr ($f>0.2$~mHz). In other words, relatively recent formation could be common, so that binaries are typically $\gtrsim10,000$~K rather than "cold" (4,000~K) by the time they pass through $f \sim 1$~mHz.} We will formulate {detailed temperature evolution from tidal heating} in the next subsection.

\subsection{Coupled orbit and temperature evolution}
\label{sec:tides}
Earlier in Figure~\ref{fig:timescale-ratio}, we classify each binary according to the strength of tidal interactions between the primary and orbit via the current instantaneous orbital decay rate in Equation~(\ref{eq:tauTF}). There we assume that the forcing frequency is simply the inverse of the orbital period. However, for the general orbital evolution in time, angular momentum must be conserved. Considering that the primary component has a rotation rate $\Omega_1$, the forcing frequency in the instantaneous tidal friction timescale Equation~(\ref{eq:tauTF}) is not $P_\mathrm{orb}^{-1}$, but instead is $f_\mathrm{orb}-\Omega_1=P_\mathrm{force}^{-1} = f/2-\Omega_1$. We present the essential equations here, but details of their derivation can be found in  Appendix~\ref{appendix:A1}.  Replacing $f/2$ with $f/2-\Omega_1$ in Equation~(\ref{eq:dfdtTD1}), the orbital frequency evolution is given by \citep{Hut1981}
\begin{equation}
\left( \frac{\mathrm{d}f}{\mathrm{d}t}\right)_\mathrm{TH} = \frac{18 k}{\mathcal{Q}} \frac{ m_2 \pi^{13/3}R_1^5 f^{13/3}\left(f/2-\Omega_1\right)}{G^{5/3} m_1(m_1+m_2)^{5/3}} .
  \label{eq:dfdtTD}
\end{equation}
As orbital energy is lost to tidal friction inside the primary, angular momentum must be conserved. Orbital angular momentum is redistributed to spin up the primary which has angular momentum $J_1 = r_g^2  m_1 R_1^2 \Omega_\mathrm{1}$. The primary rotation frequency (spin) in Hz evolves according to
\begin{equation}
 \left( \frac{\mathrm{d}\Omega_1}{\mathrm{d}t}  \right)_\mathrm{TH} = \frac{3 k}{ \mathcal{Q}} \frac{\pi^3 m_2^2 f^3 R_1^3(f/2-\Omega_1)}{ G m_1 r_g^2 (m_1 + m_2)^2}, 
\label{eq:AMev2}
\end{equation}
where the radius of gyration $r_g^2=0.1$ \citep{Hut1981}\footnote[2]{Note that \cite{Hut1981} writes this equation radian/second.}.
The total energy in the orbit and primary spin decreases from tidal heating according to 
\begin{equation}
\begin{split}
\left( \frac{\mathrm{d}E_\mathrm{tot}}{\mathrm{d}t} \right)_\mathrm{TH} &= -12 \frac{k}{\mathcal{Q} }
 \frac{ \pi^5 R_1^5{m_2}^2 f^3}{G({m_1}+{m_2})^2} \\ & \times \left( \left(f/2\right)^2- 2 \left(f/2\right) \Omega_1 + \Omega_1^2  \right).
 \end{split}
 \label{eq:Etotdot}
\end{equation}
Unlike angular momentum, which is conserved between the orbit and primary spin via Equation~(\ref{eq:AMev2}), the energy in Equation~(\ref{eq:Etotdot}) is irreversibly lost to turbulent viscosity (internal friction) in the primary. We consider that this energy loss from tidal heating is accounted for by increasing the current primary luminosity, $L_1$. Specifically, orbital energy lost by Equation~(\ref{eq:dfdtTD}) heats the primary at a rate given by Equation~(\ref{eq:Etotdot})
{and is lost through the surface through blackbody radiation}. 
{We assume that the heat conduction timescale for ELM WDs are shorter than the inspiral timescale. So we can treat WDs as being in equilibrium at all times.} 
The luminosity of the primary, or equivalently the energy loss at the surface {from radiation} assuming a blackbody is given by
\begin{equation}
    L_1 = 4\pi\sigma R_1(T_1)^2 T_1^4.
    \label{eq:lumeq}
\end{equation}

Here we pause to clarify an important point about solving for the temperature $T_1$ and radius $R_1$ using Equation~(\ref{eq:lumeq}), which is that the temperature $T_1$ cannot be determined directly by equating Equation~(\ref{eq:Etotdot}) with Equation~(\ref{eq:lumeq}).
$L_1$ is the energy loss at the surface of the finite temperature ELM WD. Until tidal interactions become important, $L_1$ is simply the residual heat since WD formation given by Equation~(\ref{eq:Lcool2}).
Once tidal interactions affect the orbit, additional energy flux from tidal heating $\left( 
{\mathrm{d}E_\mathrm{tot}}/{\mathrm{d}t} \right)_\mathrm{TH}$ will contribute to increasing the luminosity, but not completely supply $L_1$. Therefore the temperature evolution is expressed in terms of differentials, with details shown in Appendix~(\ref{sec:tempmodel}). The time evolution of the temperature can be written as
\begin{equation}
\begin{split}
& \left( \frac{\mathrm{d}T_1}{\mathrm{d}t} \right)_\mathrm{TH} = 
\frac{135 \pi^{25/3} R_1^9{m_2}^3 f^{19/3}}{G^{8/3}m_1({m_1}+{m_2})^{11/3}} \left(\frac{k}{\mathcal{Q} } \right)^2   \left(\frac{f}{2} -\frac{3}{5}\Omega_1 \right)  
 \\ &
 \times \left(\frac{f}{2} - \Omega_1\right)^2 \left(\sigma T_1^3 \left(2 R_1 + \frac{\mathrm{d}R_1}{\mathrm{d}T_1} T_1 \right)\right)^{-1},
 \end{split}
 \label{eq:dT1dt}
\end{equation}
which describes {how the surface properties of the primary respond to tidal heating.} 
\\

The sum of Equations~(\ref{eq:dfdtGW}) and (\ref{eq:dfdtTD}) for $f$, Equation~(\ref{eq:AMev2}) for $\Omega_1$ and Equation~(\ref{eq:dT1dt}) for $T_1$ form the three differential equations to solve for the temperature evolution for the primary. The radius $R_1$ appears in all three differential equations with a strong dependence (powers of 3 to 9). But the radius itself also depends on the temperature $T_1$ via Equation~(\ref{eq:RMTPanei}). In other words, the strength of tidal interactions in a binary strongly depends on radius, which in turn depends on the efficiency of the tidal heating in the primary. It is therefore a complex nonlinear problem where both the surface temperature and the radius of the primary increase as the binary is driven to Roche contact via Equations~(\ref{eq:dfdtGW}) and (\ref{eq:dfdtTD}). 

\section{Application to Galactic white dwarf binaries}
\label{sec:future-evolution}
In this Section, we consider the coupled, nonlinear time evolution of $f$, $T_1$ and $\Omega_1$ for a selection of binaries in Table~\ref{tab:binaries}. We choose the following three Galactic DWDBs: J2243, J0538 and J2029. These three binaries have the same primary mass of $0.32 M_\odot$ (see Table~\ref{tab:WD1}). However they have different companion masses, orbital frequencies (Table~\ref{tab:binaries}), and surface temperatures (Table~\ref{tab:WD1}). It will be instructive to demonstrate our model for this one single primary mass, given a variety of other initial properties.
\\

Before solving the coupled evolution with the combination of Equations~(\ref{eq:dfdtGW}) and (\ref{eq:dfdtTD}), Equation~(\ref{eq:AMev2}), and Equation~(\ref{eq:dT1dt}), we must revisit the structure and damping related constant $k/\mathcal{Q}$. Earlier in Section~\ref{sec:THmodel},
we use \cite{Piro2011}'s result that for J0651, orbital decay from tidal heating $(\mathrm{d}f/\mathrm{d}t)_\mathrm{TH}$ (Equation~(\ref{eq:dfdtTD1})) is 5.4\% of the orbital decay from gravitational waves $(\mathrm{d}f/\mathrm{d}t)_\mathrm{GW}$ (Equation~(\ref{eq:dfdtGW})). From these two Equations, we determine that the value for $k/\mathcal{Q}$ in Equation~(\ref{eq:dfdtTD1}) should be $ 8 \times 10^{-12}$ to recover that 5.4\% contribution. The ratio $k/\mathcal{Q}$ describes the ratio between $(\mathrm{d}f/\mathrm{d}t)_\mathrm{TH}$ and $(\mathrm{d}f/\mathrm{d}t)_\mathrm{GW}$. However, that expression (Equation~(\ref{eq:dfdtTD1})) for the orbital frequency increase due to tidal heating was independent of the primary rotation rate $\Omega_1$. In reality, for the general coupled evolution, an $\Omega_1$ dependence appears in the expression for the orbital decay from tidal heating (Equation~(\ref{eq:dfdtTD})). But we can still use Piro's result in the same fashion even if the quantity we solve for also depends on rotation via the dimensionless level of synchronicity $(f/2-\Omega_1)/(f/2)$, where $f/2>\Omega_1 >0$ at the initial condition $f(0)$, $\Omega_1(0)$.
\\

To rewrite Equations~(\ref{eq:dfdtTD}), (\ref{eq:AMev2}) and (\ref{eq:dT1dt}), we define the dimensionless variable $\mathcal{F}= k / \mathcal{Q}  (1- 2\Omega_1(0)/f(0))^{-1}$. Like $k/\mathcal{Q}$ in the earlier sections, this tidal contribution factor $\mathcal{F}$ is directly related to the ratio between $(\mathrm{d}f/\mathrm{d}t)_\mathrm{GW}$ and $(\mathrm{d}f/\mathrm{d}t)_\mathrm{TH}$. 
In Equations~(\ref{eq:dfdtTD}), (\ref{eq:AMev2}) and (\ref{eq:dT1dt}), $\mathcal{F}$ replaces $k/\mathcal{Q}$, so that e.g.
\begin{equation}
\left( \frac{\mathrm{d}f}{\mathrm{d}t}\right)_\mathrm{TH} = {18}{\mathcal{F}} \frac{ m_2 \pi^{13/3}R_1^5 f^{13/3}\left(f/2-\Omega_1\right)}{G^{5/3} m_1(m_1+m_2)^{5/3}} .
  \label{eq:dfdtTD3}
\end{equation}
For any $\Omega_1(0)$, $\mathcal{F}$ can be calibrated to J0651 using $ \mathcal{F} = 8 \times 10^{-12} \left(1- 2\Omega_1(0)/f(0)\right)^{-1}$ . The initial frequency $f(0)$ is the present-day gravitational wave frequency $f$ in Table~\ref{tab:binaries}, but $\Omega_1(0)$ is not known, so can be treated as a free parameter. To find $\mathcal{F}$ for a given binary to finally use Equation~(\ref{eq:dfdtTD3}) , we must choose an appropriate $\Omega_1(0)$. Since the binaries are assumed to have formed as a DWDB system some time (Myr) ago, we impose that the current coupled spin--frequency evolution is linear. In other words, we require that the ratio $2 \Omega_1/f$ is stationary for our initial $\Omega_1(0)$ and $f(0)$:
\begin{equation}
    \frac{ \mathrm{d}^2 }{\mathrm{d}f^2} \left(\frac{2 \Omega_1(0)}{f(0)} \right) = 0.
    \label{eq:initialOmega2}
\end{equation}
\\

Solving Equation~(\ref{eq:initialOmega2})
for J2243, J0538 and J2029, the initial rotation rates are 0.1mHz (5\% synchronous), 0.5mHz (44\% synchronous) and 0.28mHz (35\% synchronous) respectively.
In Table~\ref{tab:heatingmodel} we list these frequencies, along with the corresponding structure constant related product $ \mathcal{F}$ which replaces the factor $k/\mathcal{Q}$ in Equations~(\ref{eq:dfdtTD}), (\ref{eq:AMev2}) and (\ref{eq:dT1dt}). These become Equation~(\ref{eq:dfdtTD3}) for gravitational wave frequency, 
\begin{equation}
 \left( \frac{\mathrm{d}\Omega_1}{\mathrm{d}t}  \right)_\mathrm{TH} = {3 \mathcal{F}} \frac{\pi^3 m_2^2 f^3 R_1^3(f/2-\Omega_1)}{ G m_1 r_g^2 (m_1 + m_2)^2}, 
\label{eq:AMev3}
\end{equation}
for the primary spin and 
\begin{equation}
\begin{split}
& \left( \frac{\mathrm{d}T_1}{\mathrm{d}t} \right)_\mathrm{TH} = 
\frac{135 \pi^{25/3} R_1^9{m_2}^3 f^{19/3}}{G^{8/3}m_1({m_1}+{m_2})^{11/3}} \mathcal{F}^2   \left(\frac{f}{2} -\frac{3}{5}\Omega_1 \right)  
 \\ &
 \times \left(\frac{f}{2} - \Omega_1\right)^2 \left(\sigma T_1^3 \left(2 R_1 + \frac{\mathrm{d}R_1}{\mathrm{d}T_1} T_1 \right)\right)^{-1}
 \end{split}
 \label{eq:dT1dt2}
\end{equation}
for the primary temperature.
\begin{table*}
\caption{Summary of coupled primary temperature and binary evolution of the three binaries ZTF J0538, J2029 and J2243 using our tidal heating model. All three binaries have a primary mass of $0.32M_\odot$, but different companion masses of 0.45$M_\odot$, 0.30$M_\odot$ and 0.33$M_\odot$ (in listed order). The initial rotation frequency $\Omega_1(0)$ and structure constant related product $\mathcal{F} = k / \mathcal{Q}  (1- 2\Omega_1(0)/f)^{-1}$ are given by Equation~(\ref{eq:initialOmega2}), and calibrated \citep{Piro2011}'s estimate for tidal heating in the binary J0651 with Equation~(\ref{eq:dfdtTD3}). $\tau_\mathrm{RL}$ is the time until Roche contact calculated with our coupled temperature and frequency evolution model. 
Frequency $f_\mathrm{RL}$ is the binary gravitational wave frequency at Roche contact. $T_1(f_\mathrm{RL})$ is this final temperature, with $\Delta T_1$ stating the percent increase between present day temperature (Table~\ref{tab:WD1}) and Roche contact according to our model. 
{We also include two other frequencies. The first is $f_\mathrm{cold}$, which is the Roche frequency assuming the degenerate mass radius relation Equation~(\ref{eq:egg}) for the ELM WD component in each binary. The second $f_\mathrm{RL,0}$ is the Roche frequency taking the present day radius from Table~\ref{tab:WD1}, i.e. the Roche frequency if orbital decay from tidal heating is considered, only GW emission. The discrepancy between the $f_\mathrm{RL,0}$ listed here and the $f_\mathrm{RL}$ listed in Table~\ref{tab:binaries} is due to taking the radius given by the temperature dependent mass radius relation Equation~(\ref{eq:RMTPanei}) here, rather than the observed radius in Table~\ref{tab:binaries}.}
}
\label{tab:heatingmodel}
\begin{center}
\begin{tabular}{ c c c c c c c c c}
\hline
{Binary} & $\Omega_1(0)$ & $\mathcal{F}$ & $T_1(f_\mathrm{RL})$ & $\tau_\mathrm{RL}$ (kyr) & $\Delta T_1$  & {$f_\mathrm{cold}$} & {$f_\mathrm{RL,0}$}  & $f_\mathrm{RL, TH}$ 
\\

 & (mHz) & & (K) & (kyr) & (\%) & (mHz) & (mHz) & (mHz)
\\
 \hline
  ZTF J0538  &  0.5 (44\%) & $1.42 \times 10^{-11}$ &  {19,187} & {1345} & {50}  & {16} & {9.9} &  {7.7} \\
  
  ZTF J2029  & 0.28 (35\%) & $1.24 \times 10^{-11}$ & {21,728} &  {5045} & {19} & {17} & {8.2}  &  {7.3} \\
  
ZTF J2243  & 0.1 (5\%) & $8.39 \times 10^{-12}$ & {27,261} & {321} & {4}  & {17} & {6.4} &  {6.2}
\\
\hline
\end{tabular}
\end{center}
\end{table*}

\subsection{Temperature increase in J2243, J0538 and J2029}
\label{sec:future-models}
With the initial rotation frequencies $\Omega_1$ and $\mathcal{F}$ listed in Table~\ref{tab:heatingmodel}, we now model the future evolution of the three primaries in J2029, J0538, J2243 until they are no longer detached, i.e. from present day until Roche contact. In Section~\ref{sec:tides} we determined that all three are in the temperature evolution regime dominated by tidal heating (Table~\ref{tab:timescales}, Figure~\ref{fig:timescale-ratio}), with $\tau_\mathrm{TH}$ between 5 and 500 times faster than  $\tau_\mathrm{cool}$. We therefore neglect any cooling in the temperature evolution and only use the temperature increase via Equation~(\ref{eq:dT1dt2}). All three have the same primary mass $m_1=0.32~M_\odot$, but J0538 has the most massive companion $m_2=0.45~M_\odot$. 
\\

\begin{figure}
  \centering
\includegraphics[width=0.5\textwidth]{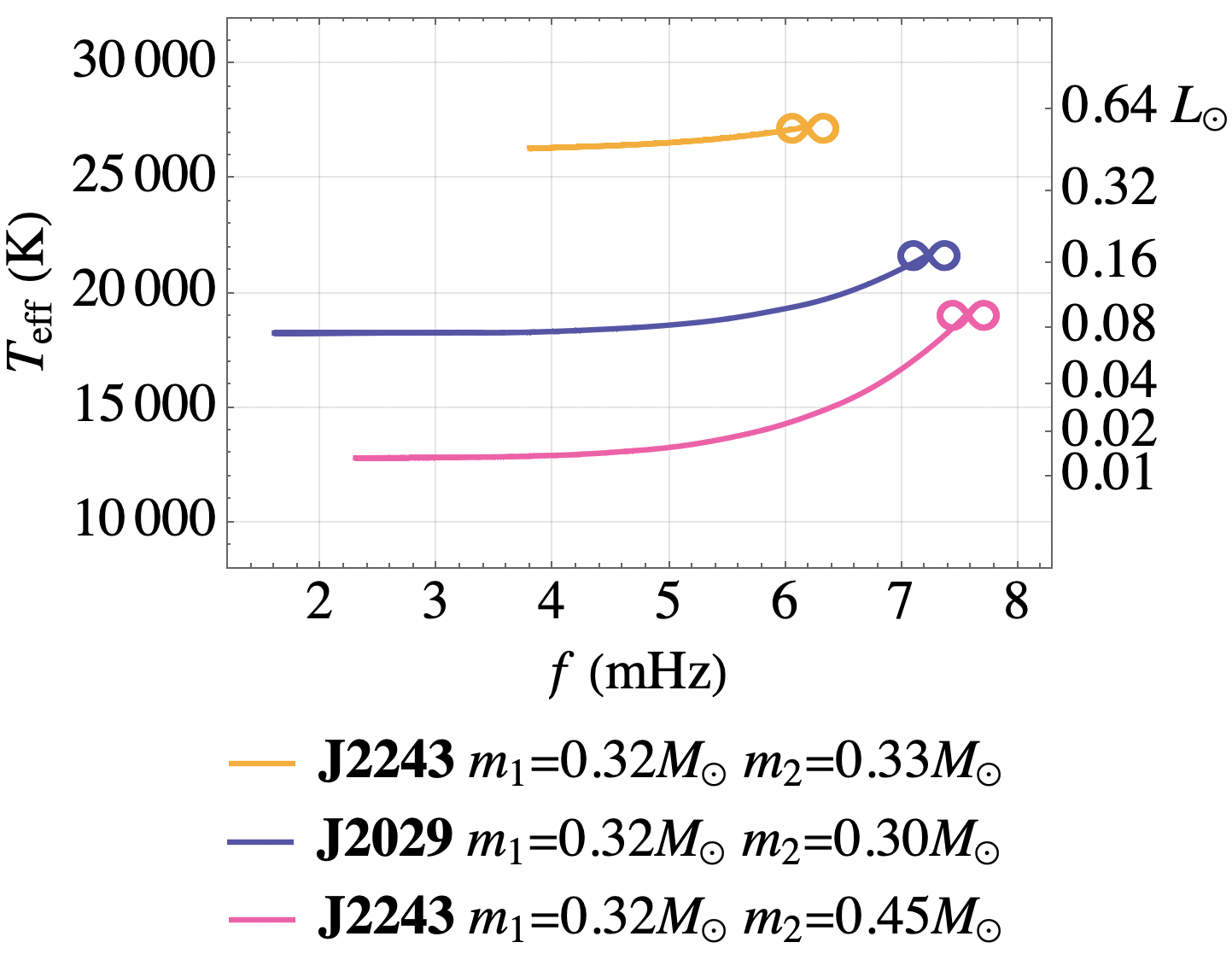}
  \caption{Evolution of temperature $T_\mathrm{eff}$ in K (left vertical axis) with gravitational wave frequency $f$ in mHz from the present day properties according to our tidal heating model for the three binaries J2243 (yellow), J2029 (indigo) and J0538 (pink). All three binaries have {the same} larger primary component of $m_1=0.32M_\odot$, {so that we can plot their luminosity on one single right hand vertical axis in solar luminosity $L_\odot$. Each of the} companion masses $m_2$ are listed in the legend. We plot each track until Roche lobe overflow {(infinity markers). This is when $f= f_\mathrm{RL}$ according to Equation~(\ref{eq:fRL}), given its temperature dependent radius from Equation~(\ref{eq:RMTPanei})}. 
  {In this temperature dependent model, the $f_\mathrm{RL}$ are 6.2, 7.3 and 7.7 mHz respectively}. However, the cold mass radius relation predicts {$f_\mathrm{RL}=16-17$~}mHz{, over two times larger.}  
  }
\label{fig:3-panel}
\end{figure}

We solve Equations~(\ref{eq:dfdtGW}) and (\ref{eq:dfdtTD3}) for $f$, Equation~(\ref{eq:AMev2}) for $\Omega_1$ {and Equation~(\ref{eq:dT1dt})  to model each binary's
frequency and temperature $T_1$ increase. We use Mathematica's ``StiffnessSwitching''
method for numerically solving ODEs. This automatically determines whether the system is stiff by ``StiffnessTest'', and switches between two extrapolation methods accordingly}\footnote[3]{These are the explicit modified midpoint (with Gragg smoothing) when stiff and linear implicit Euler when nonstiff.}. We do this for J2243 (yellow), J2029 (indigo) and J0538 (pink), shown in Figure~\ref{fig:3-panel}.
 Curves terminate at Roche contact, which is shown by the infinity markers. For the system parameters and initial conditions (starting at the left) of each curve, we take the present day $T_1$ (Table~\ref{tab:timescales}), frequency $f$ and component mass $m_2$ from Table~\ref{tab:binaries}, and the initial $\Omega_1(0)$ and structure constant shown in Table~\ref{tab:heatingmodel}. For reference, we show a scale for the luminosity in solar units $L_\odot$ in the vertical axis on the right. 
\\

Given the $m_1$ and $m_2$ in these three binaries, Roche lobe overflow occurs when $R_\mathrm{1}$ is equal to {0.38, 0.38 and 0.35} times the semi-major axis $a$ respectively. The radius $R_\mathrm{1}$ is a temperature dependent quantity (Equation~(\ref{eq:RMTPanei})), and we can see that the primary temperatures increase by {4\% (yellow), 19\% (indigo) and 50\%} (pink) in the vertical axis between present day and Roche contact. As the orbit decays via increasing $f$, the general trend is that it is easier to heat up cooler white dwarfs, suggested by these relative increases in temperature. Hotter ELM WDs have larger luminosities $\propto T_1^4$ via Equation~(\ref{eq:lumeq}). The energy dissipated in the star from Equation~(\ref{eq:Etotdot}) is radiated away at the surface, increasing in the surface luminosity. But this energy flux becomes negligible compared to the surface luminosity as the ELM WD approaches solar luminosity. Consequently the change in temperature via Equation~(\ref{eq:dT1dt}) is smaller for a hot ELM WD than a {cooler (and smaller)} ELM WD in a DWDB with the same frequency $f$.
\\

The Roche frequencies via Equation~(\ref{eq:fRL}) for these three binar{y models}, {$f_\mathrm{RL,TH}$}, depend sensitively on increasing $R_1$ and are between 6--8~mHz. These are listed in Table~\ref{tab:heatingmodel}. Before modelling the response of the temperature for the $0.32M_\odot$ primaries, the Roche frequencies {$f_\mathrm{RL,TH}$} we calculate {in Table~\ref{tab:heatingmodel} are {few--10's of percent} larger than $f_\mathrm{RL,0}$ which ignores tidal heating}. {These correspond to increases in radii of 3, 9 and 19 percent from present day.} Comparing the time from present day until Roche contact, for example, for J2243, tidal heating reduces the time until Roche contact from {350~kyr to 321~kyr (12\% sooner).}
\\

{The most important result of this section is the difference between $f_\mathrm{cold}$, and either of $f_\mathrm{RL,0}$ or $f_\mathrm{RL,TH}$ in Table~\ref{tab:heatingmodel}. This is 150--200 percent, regardless of the radius increase we predict from present day due to tidal heating until Roche contact (comparison between $f_\mathrm{RL,0}$ or $f_\mathrm{RL,TH}$). However, in population synthesis studies \citep[e.g.,][]{Nelemans2001,Marsh2004} it is standard to assume that ELM WDs are cold (Equation~(\ref{eq:egg})) at Roche contact. This results in much larger predicted Roche frequencies $f_\mathrm{cold}$ for $0.32M_\odot$ ELM WD of {16--17}~mHz. But for ELM WD with realistic temperature dependent radii (Equation~(\ref{eq:RMTPanei})), Roche contact occurs earlier (lower in frequency), demonstrated in Figure~\ref{fig:3-panel}.}
\\

Therefore, if white dwarf radii are typically 2 times larger than the cold relation at the Roche contact, then we can expect unstably mass transferring DWDB populations earlier in time, at $2^{3/2} \approx$ 3 times lower frequency $f$. For example, consider tidal heating dominated systems with modest present day $\tau_\mathrm{cool}/\tau_\mathrm{TH}$ ratios $\sim 10$ in Figure~\ref{fig:timescale-ratio}, such as J0538. When we model additional surface heating from the present day, Roche contact occurs at frequencies $\sim 10 \%$ even lower than in Table~\ref{tab:binaries}, which considered Roche contact for the temperature dependent radius but did not model future heating. Since this difference is not negligible, the Roche frequencies should be determined with this coupled model for the radius evolution, rather than simply using the present day radius for $R_1$ in Equation~(\ref{eq:fRL}). {But this difference of $\sim 10 \%$ between $f_\mathrm{RL,0}$ and $f_\mathrm{RL,TH}$ is minor compared to the $\sim 200 \%$ difference between them and $f_\mathrm{cold}$.}
\section{The He+CO double white dwarf binary J1539}
\label{sec:J1539}
In this section, we apply our model to the binary J1539 \citep{Burdge2019a} whose properties are listed in Table~\ref{tab:binaries}. So far we have been primarily concerned with the evolution of ELM WD components in detached DWDBs. Specifically, we only considered binaries whose primaries have measurements of all three of: temperature, mass and radius, and hence appear in Figure~\ref{fig:mass-radius-temp}. The ELM WD primary component of J1539 has a measured mass of 0.21$M_\odot$ and radius of $0.031R_\odot$. However it only has a temperature \textit{upper bound} of $10,000$K due to the extremely hot ($\sim$50,000~K, see Table~\ref{tab:WD1}) secondary $0.61M_\odot$ component. Therefore this binary has not been included in any of our analyses yet.   
\\

Interestingly, this temperature upper bound of $10,000$~K, shown by the yellow triangle in Figure~\ref{fig:timescale-ratio} is cooler than all other primary components with measured temperatures listed in Table~\ref{tab:WD1}. At the same time, its timescale ratio via Equations~(\ref{eq:tauTF}) and (\ref{eq:Lcool2}) is $\sim$ 1000, which is larger than all the other binaries. This large $\tau_\mathrm{cool}/\tau_\mathrm{TF}$
is due to $R_1$ being very close to the Roche radius $R_\mathrm{RL}$, so that $(R_1/a)$ is large. In section~\ref{sec:THmodel} we note {a potential} correlation between this ratio and the ELM WD temperature. {There, J1539 appears exceptionally cool given its short tidal heating timescale. Now with a formulation of tidal heating via e.g. Equation~(\ref{eq:dT1dt}), we can explain why a correlation is expected anyway, and why J1539 may be an outlier.}
\\

{At the end of Section \ref{sec:ratios} we suggest that the observed ELM WDs in Table~\ref{tab:WD1} and consequently the DWDBs in Table~\ref{tab:binaries} are relatively young $\lesssim$~Gyr. Excluding J1539's primary, the population's present day temperatures and frequencies require cooling to as low as $\gtrsim 10,000$~K after DWDB formation. Once binary frequency $f$ reaches the mHz range, GW emission brings it to Roche contact over $\sim$~Myr, so that the ELM can not cool any further. Therefore if J1539's ELM WD has a similar age $\lesssim$~Gyr then it should have had a temperature $\gtrsim 10,000$~K when $f\sim$~mHz. According to Equation~(\ref{eq:dT1dt}), temperature increases efficiently with $f$ once it is in the mHz frequency range via non-negligible $(R_1/a)^9$. Assuming detached evolution, we expect relatively high temperatures ($\gtrsim 12,000$~K) when the ratio $\gtrsim 100$ for J1539 in Figure~\ref{fig:timescale-ratio}. 
In this DWDB sample, J1539 seems exceptionally old due to its low temperature and large ratio.}  
We now explore potential formation, and past evolution of J1539 until the present day using our tidal heating model.

\subsection{Rewinding the binary evolution history of J1539}
\begin{figure}
  \centering
    \includegraphics[width=0.5\textwidth]{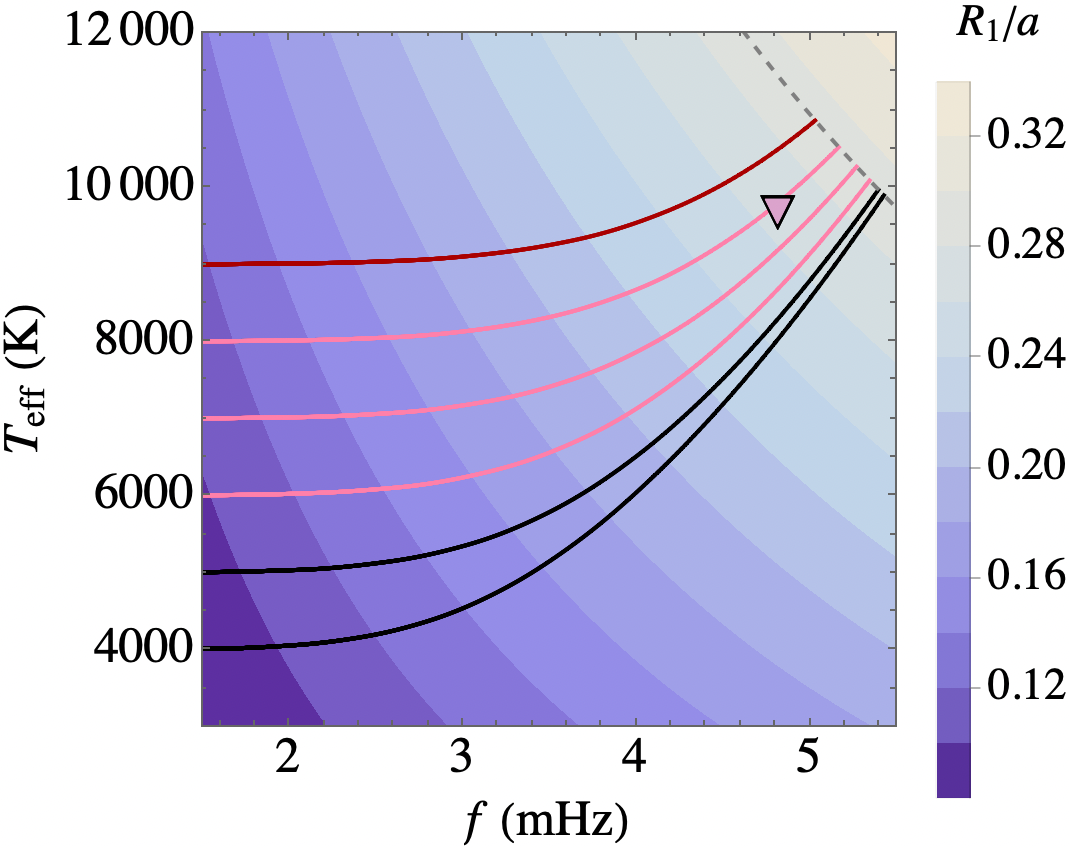}
  \caption{Temperature evolution of the primary ELM WD component in six J1539--like binaries. These models are evolved from temperatures of 4,000~K, 5,000~K, 6,000~K, 7,000~K, 8,000~K and 9,000~K at $f=1.5$ mHz, corresponding to the secondary CO WD's formation and hence DWDB formation. All tracks (solid lines) terminate at Roche contact when $R_1/a = R_\mathrm{RL}/a =$ 0.29 (grey dashed line) after around $\sim 5$ Myr. The present day gravitational wave frequency, $f=4.8$~mHz, and temperature upper bound of J1539's primary ELM WD component, $T<$10,000~K, are shown by the pink triangle. {Our tidal heating model allows the primary of a J1539 to have a temperature at the CO component formation at  $f=1.5$~mHz between 6,000-8,000~K (pink lines), and corresponding present day temperature between 8,000--10,000~K. We ultimately exclude the three black and maroon curves (see text for details).}
  }

\label{fig:J1539}
\end{figure}

We investigate several models for the prior frequency and temperature evolution of the  0.21$M_\odot$ ELM WD primary in J1539.
With these tidal heating models we attempt to rewind the binary's evolution, given its present day properties. Depending on which models comply with the current upper limit of the temperature {(pink triangle marker in Figure~\ref{fig:J1539})}, we can potentially place constraints on the binary evolution history of J1539. We take a ``J1539--like'' $0.21+0.61M_\odot$ detached binary, and consider a variety of evolution histories associated with double white dwarf binary (DWDB) formation.
We define DWDB formation as the moment the second white dwarf forms. Typically, in a hybrid He+CO system this is thought to be the lighter Helium component \citep{Iben1993}. However, according to Equation~(\ref{eq:Lcool2}) the hot CO component must have formed only 5.35 Myr ago. Assuming that the binary is detached, since the He primary is at least hundreds of Myr old, the CO component formed second. Then, the detached DWDB system has been decaying via Equation~(\ref{eq:dfdtGW}) for the past 5.35 Myr. The present day frequency $f$ is 4.8mHz, corresponding to a Roche contact time in {$30$~kyr} (Table~\ref{tab:binaries}). Solving Equation~(\ref{eq:dfdtGW}) back in time until DWDB formation 5.35 Myr ago gives a DWDB formation frequency of 1.5 mHz. Therefore we consider $f=1.5$ mHz (22 minute orbital period) to be the initial binary frequency our J1539--like binary models. Cooling, which occurs over Gyr can be completely ignored since a J1539--like binary at this frequency of $f=1.5$ mHz will still reach mass transfer relatively quickly, in $\sim 5$ Myr. 
\\

Now we consider the temperature increase in the 0.21$M_\odot$ ELM WD primary from $f=1.5$mHz until Roche contact, for six temperatures at DWDB formation: 4,000~K, 5,000~K, 6,000~K, 7,000~K, 8,000~K and 9,000~K. These six temperatures for the ELM WD at CO WD formation at $f=1.5$mHz are at the left-hand boundary for each solid line in Figure~\ref{fig:J1539}. 
Here, we used that the initial rotation rate $\Omega_1$ is almost synchronous with the orbit ($1 - 2 \Omega_1/f = 10^{-4}$). Each binary evolution model terminates at its respective intersection with the grey dashed line, which is when the primary radius $R_1 = R_\mathrm{RL} =$ 0.29$a$ via Equation~(\ref{eq:RRL}). The Roche frequencies $f_\mathrm{RL}$ for each J1539--like binary are determined by Equation~(\ref{eq:fRL}), and are listed in Table~\ref{tab:J1539}. They range from 5.0 mHz (9,000~K initial temperature) to 5.4 mHz (4,000~K). By comparing the solid lines' intersection with the grey dashed line in Figure~\ref{fig:J1539}, we can see that the Roche frequencies are smaller for hotter, hence larger primary components. 
\begin{table}
\caption{Formation and Roche contact related properties in models of the binary J1539 for various assumptions about the 0.21$M_\odot$ primary ELM WD component. Assuming that the CO white dwarf formed second at $f=1.5$ mHz, we consider that the older ELM WD component had temperatures ranging from 4,000--9,000~K there. We list the temperatures predicted by the model for each track at present day (4.8~mHz), which range from around 8,000--10,500~K.
}
\label{tab:J1539}
\begin{center}
\begin{tabular}{ c c c c}
\hline
$T(1.5$) (K) & $f_\mathrm{RL}$(mHz) & $\tau_\mathrm{delay}$(Gyr) & $T(4.8$) (K) \\
 \hline
  4,000  & 5.43  & 5.6 &  7986   \\
  
  5,000  & 5.40 & 2.3 &  8253  \\
    
  6,000  & 5.34 & 1.1 &  8643 \\
    
  7,000  & 5.27 & 0.58 &  9153   \\
    
  8,000  & 5.17 & 0.33 &  9767  \\
    
  9,000 & 5.03 & 0.20  & 10467   \\
  \hline
\end{tabular}
\end{center}
\end{table}
In all cases, the Roche contact time $\tau_\mathrm{RL}$, which is the time taken for curves to reach the dashed grey line via Equation~(\ref{eq:tRL}) is around 5 Myr. 
\\

Using Figure~\ref{fig:J1539}, we can now speculate about possible evolutionary histories of J1539. The temperature of the primary is $T_1<10,000$~K. However the evolution of the hottest initial condition (9,000 K, upper most track in maroon) leads to a present day temperature of around 10,500~K. Our tidal heating model therefore excludes temperatures at 1.5mHz to be $\lesssim 9,000$~K. Additionally, we recall that in this scenario that we are considering, the ELM WD primary formed first. It has therefore had time to cool from its own white dwarf formation, until when the secondary finally forms a white dwarf at 1.5~mHz. This was 5.35~Myr ago, at DWDB formation. Using Equation~(\ref{eq:Lcool2}), we estimate the delay time $\tau_\mathrm{delay}$ since the ELM WD primary's formation, which should be much longer than 5.35~Myr ago, until all six temperatures at $f=1.5$~mHz. In order of increasing temperature at $f=1.5$~mHz, these are 5.6, 2.3, 1.1, 0.58, 0.33 and 0.20 Gyr ago, respectively, and are listed in Table~\ref{tab:J1539}. 
\\

Based on initial -- final mass relations for white dwarfs \citep{ElBadry2018,Hollands2024}, the progenitor of the 0.61$M_\odot$ secondary had a ZAMS mass between $1.6-2.2M_\odot$. This corresponds to lifetimes for the progenitor of around 1~Gyr. 
If we assume that the CO WD is very young as suggested by its temperature, then the lifetime of its progenitor star sets the upper limit on the binary's age for the required binary evolution to make the He+CO DWDB. Then the ELM WD primary's progenitor must have formed a white dwarf within the last 1~Gyr. The two coolest initial conditions (4,000~K and 5,000~K, black solid lines) require longer cooling times (5.6 and 2.3~Gyr) than the CO progenitor's lifetime, presenting a contradiction. Assuming that J1539 is detached, and that the component temperatures can only be explained by white dwarf cooling and tidal heating, only the solid pink curves starting at 6,000~K, 7,000~K and 8,000~K are allowed according to our model.
\\

In \cite{Burdge2019b}, the authors speculated that the unconventional component temperatures (cold He + hot CO component) may be a sign that the binary has already started mass transferring. If J1539 has already reached Roche lobe overflow, then the He donor can be cool and the secondary hot because of accretion. Our model is only suitable for detached binaries, whose cooling of residual temperature and tidal heating are modelled by Equations~(\ref{eq:Lcool2}) and~(\ref{eq:dT1dt}). Therefore, if future follow up observations suggest that the present day temperature of the primary is $<5,000$~K, then according to our tidal model, J1539 is not detached. Or if mass accretion onto the CO component can be confirmed with e.g. X-ray observations, then this could provide an explanation for the ELM primary J1539 appearing as an outlier in Figure~\ref{fig:timescale-ratio}.

\section{Discussion}
\label{sec:discussion}
\subsection{The observed double white dwarf population with eclipsing methods}
Here we discuss subtleties and speculate about calculating the number distribution $N(f)$ of short period DWDBs given by Equation~(\ref{eq:NZTF}). There, the intrinsic number density in gravitational wave frequency $f$ space $N(>f) \propto f^{-8/3}$. This ignores the chirp mass distribution, but we note that it seems sharply peaked at $0.3M_\odot$ in Table~\ref{tab:binaries}. Then for characterising the electromagnetic (EM) selection effects, the eclipsing probability of a binary is $\left( R_1+R_2 \right)/a \propto \left( R_1+R_2 \right) f^{2/3} (m_1 + m_2)^{-1/3}$. Ignoring the mass ratio dependence, and radius of the primary, the detection probability by eclipsing scales by a factor of $R_1 f^{2/3}$. Assuming a fixed $R_1$ for each binary during its time in the mHz band (no tidal heating), the number distribution in EM becomes $N(>f) \propto f^{-2}$. 
We expect the eclipsing white dwarf population to follow this power law, where for large enough $N$ we expect to see 9 times as many binaries at 1~mHz compared to 3~mHz. It is therefore surprising that the eclipsing binary population of nine ELM DWDBs in Table~\ref{tab:binaries} may be uniformly populated between {1--5~mHz (range of $f$ in Table~\ref{tab:binaries}). This should make the dashed parts of the observed population $N_\mathrm{ZTF}$ in Figure~\ref{fig:NZTF} flatter. In fact, this apparent dearth of lower frequency binaries with J0651--like ELM WD temperatures was pointed out in \cite{Piro2011}, when the sample size was much smaller.}
\\

In Figure~\ref{fig:3-panel} and \ref{fig:J1539} it is clear that tidal heating generally is more efficient for larger frequency $f$. Equivalently, tidal heating is more efficient with larger eclipsing probabilities, since $f\propto a^{-3/2}$. This is because of two interdependent effects. First, the radius $R_1$ for each binary increases during the mHz band according to Equation~(\ref{eq:RMTPanei}) as it is heated by Equation~(\ref{eq:dT1dt}). Second, the rate of heating by Equation~(\ref{eq:dT1dt}) increases with larger $f$ anyway, since the separation $a$ is smaller. This additional selection effect of larger average $R_1$ with increasing $f$ should be included in $N(f)$, since larger $R_1$ results in larger luminosities. In this case, the observable volume could be frequency dependent, on average. For example, by employing our tidal heating model in a simulated Milky Way DWDB population (mass ratios, component masses, and binary formation frequencies), the $R_1$ dependence on $f$ can be estimated.
\\

With this, it may be possible to estimate how the detectability of DWDBs via EM ``selection'' in Equation~(\ref{eq:NMW0}) increases with $f$ on average in a given Galactic population. Enhanced detectability at larger $f$ will be due to both the eclipsing factor and observable volume via the luminosity increasing with $R_1$. For now, we speculate that there could be a dearth in binaries detected by eclipsing around 1mHz, due to not being tidally heated yet, compared to higher frequencies. We suggested in Section~\ref{sec:RLfreq} that due to the tidal heating effects, the lightest primaries in the population will start mass transfer around 2mHz. Even so, the frequency dependence of the power law distribution $N(>f)$ may become flatter than $f^{-2}$ after around 3mHz, when ELM WD with primaries $\gtrsim 0.2M_\odot$ are efficiently tidally heated to increase their luminosity (see Figures~\ref{fig:3-panel} and \ref{fig:J1539}). This prediction for the dependence on $f$ can be tested once there is a sufficiently large population (exceeding $\sim$ hundreds) of DWDBs detected in EM.

\subsection{Comparison with previous work}
Tidal heating in the short-period double white dwarf binary population has previously been modeled by \cite{Scherbak2024}.
The objective of that work was to estimate formation frequencies of known DWDBs, and constrain common envelope processes. However, we can make a qualitative comparison between our models and their models. 
The strength of dynamical g--mode excitation and heating rate {\citep{Burkart2013,Fuller2011}} depends on the details of the g--mode frequency spectrum in a specific stellar model, and is quantified with a dimensionless amplitude $F(\omega)$, which depends sensitively on the forcing frequency $\omega$. In the context of our equilibrium tide framework, $F(\omega)$ can be thought of in analogy with the coupling and damping rate related constant $k/\mathcal{Q}$ in Equation~(\ref{eq:tauTF}), where $F(\omega)=3k_2/(2\mathcal{Q})$.
We used a frequency independent $k/\mathcal{Q}$ corresponding to the coupling and damping for the quadrupolar f--mode. This mode has the strongest coupling via $k$, but orbital resonances do not play a role due to the relatively high frequency of f--modes in WD ($\sim 100$~mHz).
On the other hand, g--modes, which typically have frequencies $\omega \sim$~mHz, are weakly coupled to the orbit via $k$ \citep[e.g.,][]{Lee1986}. However, 
when the current orbital frequency $\omega$ (forcing frequency) is close to resonance with a g--mode, the frequency dependent amplitude $F(\omega)$ becomes several orders of magnitude larger. Then strong dissipation can occur comparable with the strongly coupled quadrupolar f--mode from the frequency independent framework. Therefore in models of tidal heating in detached binaries until Roche contact, we expect similar temperature increases.
\\

When the detached DWDBs shown in Figures~\ref{fig:3-panel} and \ref{fig:J1539} reach $f\sim$~few~mHz, we find the same qualitative trend of a runaway temperature $T_1$ increase with {the dynamical tide models of \cite{Scherbak2024}}. Here, $T_1$ increases $\sim 10$~\% between when $f \gtrsim$~few~mHz and Roche contact. Although there is no direct comparison between our Figures, it seems that we generally predict slightly less (factor 2) temperature increase than theirs. This could be due to the coupling being over predicted via the fit for $F(\omega)$ in {\cite{Burkart2013}} and \cite{Fuller2013} {where the low coupling between resonances is not captured}. Another reason is that our model calibrated the coupling and damping $k/\mathcal{Q}$ to observations \citep{Piro2011}, whereas \cite{Scherbak2024} calculated this directly from specific stellar models. In any case, the overall similarity is reassuring. 
\\

{With an equilibrium tidal heating model, we assume that the dissipated energy is conducted throughout the white dwarf over a timescale much faster than the inspiral timescale, like in \cite{DallOsso2014}. For the $0.3M_\odot$ primaries in J2243, J2029 and J0538 (Figure~\ref{fig:3-panel}), we calculated change in radii of 3, 9 and 19 percent. Although we do not model J0651, our predicted increases $\lesssim 10\%$ seem compatible with the predicted radius increase $\sim 5 \%$ of \cite{DallOsso2014}.
Once the inspiral timescale becomes shorter than the heat conduction timescale in the white dwarf, it may be inappropriate to use the equilibrium models of \citep{Panei2000} for the radius evolution. In this case, more detailed stellar structure models are required to calculate the radius according to the location of energy deposition.
}
\\

{By taking a universal $k/\mathcal{Q}$ for ELM WDs, the model presented in this paper described by Equations~(\ref{eq:dfdtGW}) + (\ref{eq:dfdtTD}), (\ref{eq:AMev2}) and (\ref{eq:dT1dt}) generalizes the coupled orbit and ELM WD temperature evolution. It is therefore ready to apply to e.g. simulations of DWDB populations in the Milky Way. It can be modified if, for example, the appropriate universal $k/\mathcal{Q}$ value is revised with future ELM DWDB observations. If $k/\mathcal{Q}$ is significantly larger (i.e. much smaller tidal friction timescale), for a given binary with $R_1/a$ the $\mathrm{d}f/\mathrm{d}t$ from tides is much larger. But to quantify the final temperature at Roche contact, the coupled model must be used because of the interdependent evolution of frequency and temperature. On the other hand, $k/\mathcal{Q}$ can only be a factor of a few times smaller than what was used in this work, given the upper limit for $\mathcal{Q}$ suggested by \cite{Piro2011}. Even in this limiting case, we expect milder but still non negligible temperature and radii increases than Figure~\ref{fig:3-panel}.}

\subsection{{The Galactic double white dwarf population detected in gravitational waves}}
\label{sec:population_discussion}
Finally, we comment on the implications of our results for the detection of Galactic white dwarf binaries by space-based gravitational wave detectors such as LISA \citep{LISA2017}, Taiji \citep{Taiji2017}, and TianQin \citep{Tianqin2016}. These missions are capable of detecting $\gtrsim 10^4$ double white dwarf binaries in the Milky Way which have orbital periods $<$1 hr. The detection of DWDBs in gravitational waves requires a correct interpretation of their measured $f$ and $\dot{f}$. For detached DWDBs, $\dot{f}$ is given by Equations~(\ref{eq:dfdtGW}) + (\ref{eq:dfdtTD}), so that chirp mass estimates should be relatively straightforward. However, if a white dwarf component in a given binary is mass transferring, its $\dot{f}$ depends on the stability of mass transfer \citep{Soberman1997}. An ELM WD binary undergoing unstable mass transfer will merge on the dynamical timescale, where the positive frequency increase $\dot{f}>\dot{f}_\mathrm{GW}$. But for a stably mass transferring ELM WD binary, the frequency derivative eventually becomes negative, $\dot{f}<0$. 
\\

To interpret DWDB populations detected by gravitational waves, the steady state assumption used in Equation~(\ref{eq:NMW}) is valid so long as the population is not subject to unstable mass transfer \citep{Seto2022}. 
 Assuming the cold relation for $R_1$ given by Equation~(\ref{eq:egg}), it is valid until the Roche frequency for the lightest possible ELM WD primary mass of 0.15$M_\odot$. This is $f<7$ mHz via Equation~(\ref{eq:fRL}).  
In this paper we propose that tidal heating generically results in helium composition white dwarfs which are $\gtrsim 10,000$~K at Roche contact, or at least 1.5 times larger than the cold relation. In other words, we expect Galactic ELM WD in DWBDs to be large and hot at the onset of mass transfer. In this case, then via Equation~(\ref{eq:RMTPanei}), Roche frequencies should be smaller according to $f_\mathrm{RL} \propto R_1^{-3/2}$. In Figure~\ref{fig:mass-radius-temp}, the lightest primary component is J0533's $0.167~M_\odot$ primary, which is 20,000~K. Using Equation~(\ref{eq:fRL}) with its actual inflated radius, the onset of mass transfer is expected at $f_\mathrm{RL}=1.9$~mHz. If we consider the lightest possible 0.15~$M_\odot$ primaries to typically have radii twice the cold relation, corresponding to 12,000~K, then $f_\mathrm{RL}=2.4$~mHz. Our results therefore suggest that the population's frequency evolution may be affected by mergers from around 2~mHz, due to larger white dwarf radii inflated by tidal heating. This is several times smaller than the estimated 7 mHz. Caution should therefore be taken when interpreting any resolved candidate DWDB gravitational wave source with a frequency $f\gtrsim2$~mHz to estimate its chirp mass via Equation~(\ref{eq:dfdtGW}).
\\

Earlier Roche contact frequencies in the DWDB population has implications for the unresolved DWDB ``foreground'', which is expected to hinder the detection of individual binaries in e.g. LISA \citep{Ruiter2010,Cornish2017}. Estimates of the foreground generally drop off very steeply to disappear by a few mHz. But these estimates depend sensitively on assumptions related to typical mass transfer stability, which assumes cold white dwarfs. For example, recent work by \cite{Toubiana2024} found that predictions for the confusion noise between 1--3~mHz can differ by up to a factor 2 depending on assumptions about angular momentum transport during stable mass transfer. In this work we find that in the Galactic DWDB population, after around 2~mHz, the scaling of frequency with the number density $N(f) \propto f^{-8/3}$ no longer holds. This is due to the binaries in the population (with the lowest mass primaries) disappearing due to merging from 2~mHz. But for the cold relation, DWDBs are expected to start disappearing at 7~mHz, which is much higher than the range of frequencies where substantial confusion noise is expected, 1--3~mHz. We speculate that our finite temperature treatment, specifically its consequence on the typical Roche contact frequencies should change the shape of confusion noise estimates for a given population.

{\subsection{Mass radius relation for ELM WD in DWDBs}}
{
In this paper, we took \cite{Panei2000}'s He WD models which have hydrogen envelopes, $3 \times 10^{-4}M_\odot$. However they also presented mass radius relations for He WD without a H envelope. At most, for the $0.15-0.5M_\odot$ mass range investigated these can be up to 30 percent smaller than the H envelope models for a given mass. Without the H envelope, the radius depends more sensitively on the surface temperature:
$\mathrm{d \ ln}R/\mathrm{d \ ln}m \propto -T_\mathrm{eff}^{3/2}$. Assuming this mass radius relation without the hydrogen envelope changes our key point about the onset of mass transfer. Rather than occurring as early as $f=2$~mHz, it can occur only from higher frequency of 3.4 mHz due to more compact radii. However, we find that observations fit better with the H envelope models for Figure~\ref{fig:mass-radius-temp}. In addition, observed ELM DWDBs have some amount of hydrogen leading to their spectral classification as double-lined DA binaries (except J0533). We do not expect pure helium in detached systems anyway since complete removal would require some mass transfer history of the larger ELM WD.}
\\

{In addition, we modelled all primaries as having He composition. However, several studies \citep[e.g,][]{Moroni2009,Scherback2023} suggest that CO composition white dwarfs can have masses as low as 0.32$M_\odot$. For these low mass CO WDs, mass radius relations can be seen in \cite{Panei2000}'s Figure 4. At around $0.3 M_\odot$, radii are 
$\sim$~10--20 percent smaller than a He composition WD with the same mass. Therefore if DWDBs containing ELM WD are detected by gravitational waves only, this possibility of more compact radii due to CO composition should be considered in conjunction with the initial final mass function for WDs.}

\section{Summary and Conclusions}
\label{sec:sec6}
We presented a model for tidal heating in short period double white dwarf binaries, based on the formulation of tidal friction in \citet{Hut1981}. 
Using a temperature-dependent mass--radius relation for helium composition ELM WDs based on \cite{Panei2000}, we modeled the coupled frequency and temperature evolution in several Galactic binaries. These binaries have primary ELM WD components with surface temperatures between 12,800--26,300~K, and their orbital separations will decay over k--Myr timescales primarily from the emission of gravitational waves. We predicted that their surface temperatures can be heated from their present day temperature by tens of percent until Roche contact. 
\\

We also present a set of models for tidal heating in a J1539--like binary with a $0.21$~$M_\odot$ He primary and a $0.61$~$M_\odot$ CO secondary. We explore possible prior evolutionary tracks related to J1539's DWDB formation and the cooling of the He primary. There it is assumed that the hot CO component formed second according to its present day temperature 5.35~Myr ago. For relatively cool temperatures between 4--9,000~K at $f=1.5$~mHz (5.35~Myr ago) we predict surface temperatures at Roche contact around 5~mHz between 9--11,000~K.
 For example, a degenerate 4,000~K (when $f=1.5$~mHz)
 ELM WD component in a decaying binary will double its temperature to be 8,000~K at Roche contact. We constrain the past evolution of J1539's ELM WD primary to be 6,000--8,000~K when the binary was 1.5~mHz, and have an age between around 0.3 and 1.1~Gyr. Alternatively, the unexpected temperatures (hot CO component, cool He component) may simply be because J1539 is mass transferring and not detached. In this case, our model cannot be used to constrain its prior {evolution.}
\\

In our coupled temperature and binary frequency evolution model for ELM WD in DWDBs, the {fractional increase in temperature} is more pronounced in initially cooler white dwarfs. This is simply because it is more difficult to heat up an already luminous WD given some additional energy flux, which in this case is sourced by tidal interactions and internal friction.
As the ELM WD is heated, its 
luminosity increases towards stellar luminosities (see Figure~\ref{fig:3-panel}), {raising the surface temperature}. This temperature increase is responsible for lifting the degeneracy, where the outer layers become thermally supported.
The current short period binary population listed in Tables~\ref{tab:WD1} and \ref{tab:binaries} typically have $R_1/a\sim 0.1$, with $T_1\gtrsim 20,000$~K. According to our Figure~\ref{fig:timescale-ratio} many are only in the mildy tidally heated regime, since they are still k--Myr away from Roche contact. If we assume that these binaries formed $\sim$Gyr ago, this tidal heating model cannot produce their temperatures. Based on cooling timescales, these binaries likely have lifetimes $\lesssim 100$ Myr, so that their temperatures are better explained with residual heat. In other words, we cannot make the observed population from completely cooled ($T_1 \sim 4,000$~K) initial conditions.
\\

However for DWDB systems close to Roche contact, if their primary temperatures are $T_1\lesssim$
12,000~K, the following scenario is possible: the binary is very old and formed as a DWDB $\sim$10~Gyr ago, its lighter ELM WD component completely cooled to around 4,000~K, and then was heated up again by inevitable tidal heating once the orbit decays and $R_1/a\gtrsim 0.1$. Even for these very old binaries, at the time of Roche contact the primary should still be at least $\sim 10,000$~K according to our model. In addition, the radii of ELM WDs at Roche contact should always be at least 1.5 times the degenerate relation (Figure~\ref{fig:mass-radius-temp}) at Roche contact. The lower mass ELM WDs between $\sim$0.15--0.25~$M_\odot$ should be at least twice as large. This corresponds to Roche lobe frequencies of $f_\mathrm{RL}=2$~mHz, which is over three times lower than the Roche lobe frequency calculated assuming that ELM WDs are degenerate (7~mHz).
\\

This has direct implications for proposed space-based gravitational wave detectors, which are expected to detect at least $\sim10^4$ short period DWDBs. In the population, the lightest possible ELM WDs will have masses $\sim0.15~M_\odot$. In general, ignoring details of temperature dependence, these will likely be the largest ELM WDs. Assuming the degenerate mass--radius relation, an ELM WD with $m_1 \sim0.15~M_\odot$ begins donating mass at around 7~mHz according to Equations~(\ref{eq:egg}) and (\ref{eq:fRL}). However, we expect these to have radii twice as large as the degenerate relation by the time of Roche contact. Then, ELM WD donors with mass $\sim0.15~M_\odot$ reach Roche lobe overflow much earlier, at 2~mHz. Therefore the DWDB population begins to be affected by unstable mass transfer leading to mergers around 2~mHz, not 7~mHz. Therefore for individually resolved binaries with frequency $f\gtrsim2$~mHz, great caution should be taken in interpreting the $\dot{f}$ to relate directly to the chirp mass via Equation~(\ref{eq:dfdtGW}). 
\\

This also has consequences for the unresolved DWDB population. The degenerate assumption in Equation~(\ref{eq:egg}) is standard for constructing the confusion-limited foreground \citep{Ruiter2010}. For these estimates, white dwarfs are usually evolved until Roche contact according to Equations~(\ref{eq:egg}) and (\ref{eq:fRL}), after which they either quickly merge, or survive as stably mass transferring DWDB. {If the observed sample is representative of the Galactic intrinsic DWDB population, then the population $N_\mathrm{MW}$ (Figure~\ref{fig:NZTF}) is reduces further with $f$ due to mergers after $f\gtrsim2$~mHz. On the other hand, for cold ELM WDs the frequency distribution of the population can be assumed to be given by Equation~(\ref{eq:dfdtGW}) until 7~mHz.} It is unclear whether this {lower frequency for the onset of mass transfer in the DWDB population} enhances or reduces confusion noise in the $f=$ 1--3~mHz range. 
{But our general tidal heating model for the coupled ELM WD temperature and binary frequency evolution can be readily utilized in already simulated populations of DWDBs. By determining the properties at Roche contact of simulated Galactic DWDB populations, more realistic confusion noise curves which account for generically hot ELM WD can be calculated.}
\\

{For DWDBs, binaries with critical mass ratios $q>2/3=q_\mathrm{crit}$ are dynamically unstable to mass transfer and will merge \citep{Soberman1997}.} {This standard criteria is based on the degenerate mass radius relation. Then, if various modes of mass transfer are considered \citep[e.g.,][]{NelemansMT,Marsh2004}, the various criteria suggest $q_\mathrm{crit}<2/3$, i.e. unstable mass transfer can happen for lower mass ratios. In any case, all of these} criteria {rely} on the assumption that the WD donors are completely degenerate when they start mass transferring at Roche contact. 
\\

{In this paper, we suggest} that ELM WD are typically hot and large at the onset of mass transfer{. This is due to a combination of: DWDB formation frequencies $\gtrsim 0.1$~mHz which do not allow the ELM WD components to completely cool, and also inevitable tidal heating when $f\gtrsim 2$~mHz.} The consequences of this on the physics and stability of mass transfer of ELM WDs in DWBDs was not in the scope of this work. So a natural next step is to consider the finite temperature dependent mass transfer stability of white dwarfs. 
\\

{From our results, we can already speculate that finite temperature ELM WD in DWDBs should be more stable to mass transfer compared to \cite{Soberman1997}'s $q>2/3$ criteria for a DWDB merger.} This is because they should expand \textit{less} than their completely degenerate counterparts, due to being partially thermally supported in the outer layers. {This may also work to stabilize the criteria which account for the physics of mass transfer and tidal torques which act to destabilize the onset of mass transfer \citep{NelemansMT,Marsh2004}. }However, detailed stellar models \citep[e.g.,][]{Wong2021} will be required to formulate a stability criteria which depends on a given ELM WD's thermal properties and details of mass transfer. 
\\

Our predictions about large radii and hot temperatures at Roche contact {also} has several implications for transient phenomena and stellar populations detected by electromagnetic telescopes. In particular, He+CO white dwarf binaries which reach Roche contact are candidate progenitors for Type .Ia supernovae explosions, stably mass transferring AM CVn binary systems, and RCrB stars. The precise branching of these three scenarios, and hence expected rates of each outcomes depends not only on the stability of mass transfer, but also on the nuclear stability of the accreting CO white dwarf. For example, formation of an RCrB star requires unstable mass transfer without any thermonuclear explosion.
\\

In this paper, we suggest that ELM WD donors in the progenitors of such exotica should generically be at least $10,000$~K {at the onset of mass transfer}, and therefore have radii at least 1.5 times the degenerate radius for WD. {This is the case even if our equilibrium tides model for tidal heating (e.g. radius response) is found to be not appropriate, due to $<$~Myr binary inspiral timescales of the observed DWDB population. }
This should be taken into account not only for future gravitational wave detection, but also transient rates and stellar populations detectable by {currently operating} electromagnetic telescopes. 
\typeout{}

\section*{Software and third party data repository citations}
 Mathematica notebooks to reproduce all Figures in this work are available at \href{https://github.com/mcneilllucy/mathematica}{https://github.com/mcneilllucy/mathematica}. 

\begin{acknowledgements}
LM is grateful to Naoki Seto and Jim Fuller for helpful discussions, encouragement, and valuable feedback. {We thank the referee for their careful reading, invaluable insights and thoughtful suggestions which greatly improved this paper.}
LM acknowledges financial support from the Japan Society for the Promotion of Science (JSPS) International Research Fellow program (Graduate School of Science, Kyoto University JSPS P21017), RIKEN iTHEMS and the Hakubi project at Kyoto University. \textsc{Mathematica} was used to perform calculations and generate figures in this work.
\end{acknowledgements}

\software{Wolfram Mathematica \\  (https://www.wolfram.com/mathematica)}

\newpage

\appendix
\section{Response of the surface temperature and rotation}
\label{appendix:A1}
For general $\Omega_1>0$, the forcing period $P_\mathrm{force}$ in Equation~(\ref{eq:tauTF}) is inverse of the forcing frequency via $f_\mathrm{orb}-\Omega_1=P_\mathrm{force}^{-1} = f/2-\Omega_1$. Then the evolution of the gravitational wave frequency $f$ of a circular binary due to tidal heating is given by {Equation~(\ref{eq:dfdtTD}) which we rewrite here}:

\begin{equation}
\left( \frac{\mathrm{d}f}{\mathrm{d}t}\right)_\mathrm{TH} = \frac{18 k}{\mathcal{Q}} \frac{ m_2 \pi^{13/3}R_1^5 f^{13/3}\left(f/2-\Omega_1\right)}{G^{5/3} m_1(m_1+m_2)^{5/3}} .
  \label{eq:dfdtTD0}
\end{equation}
{Equation~(\ref{eq:dfdtTD0})} has the following limits: far from synchronous when $\Omega_1 \to 0$, this reduces to the non rotating Equation~(\ref{eq:dfdtTD1}). When the primary spin and orbit are close to synchronous, to first order in $f/2-\Omega_1$ the frequency decay is given by

\begin{equation}
  \left( \frac{\mathrm{d}f}{\mathrm{d}t}\right)_{\mathrm{TH,}f/2\to\Omega_1} =  \frac{k}{\mathcal{Q}} \frac{ 144 \times 2^{1/3} \Omega_1^{13/3} m_2 \pi^{10/3}R_1^5}{G^{5/3} m_1(m_1+m_2)^{5/3}}\left(f/2-\Omega_1\right).
\end{equation}
As required, in this constant lag time model the decay from tidal heating via Equation~(\ref{eq:dfdtTD}) is zero for the synchronous state $f=2 {\Omega_1}$ outlined in \cite{Hut1980}. However for double white dwarf binaries containing an ELM WD component, the frequency decay {is dominated by energy loss from gravitational wave emission Equation~(\ref{eq:dfdtGW}). The tidal contribution is at most $\sim 5$ times the gravitational wave contribution. Therefore a synchronous state cannot be assumed,} and we must consider the spin evolution for a general $\Omega_1$, given changing $R_1${and also $f$.}
\\

The binary orbital angular momentum for $m_1$, $m_2$ separated by semi-major axis $a$ in terms of the gravitational wave frequency $f$ is
\begin{equation}
    h(t) = \frac{G^{2/3} m_1 m_2}{f(t)^{1/3}(m_1+m_2)^{1/3} \pi^{1/3}} \propto a(t)^{1/2}.
    \label{eq:h-f}
\end{equation}
The primary with subsynchronous rotation $\Omega_1$ is {spun up through tidal torques}, due to binary angular momentum loss described in {Equation~(\ref{eq:dfdtTD0})}.
The primary's angular momentum is given by $J_1(t) = I_1 \Omega_\mathrm{1}(t)$, and the moment of inertia $I_1 = r_g^2 m_1 R_1^2$, where $r_g^2=0.1$ \citep{Hut1981}. By conservation of angular momentum the evolution of the primary spin $\dot{J}_1(t)=-\dot{h}(t)$ (Equation~(\ref{eq:h-f})). The change in primary angular momentum $J_1$ from tidal friction can therefore be written:
\begin{equation}
         \left( \frac{\mathrm{d}h}{\mathrm{d}t}  \right)_\mathrm{TH}  = -\left( \frac{\mathrm{d}J_1}{\mathrm{d}t}  \right)_\mathrm{TH}= \left( \frac{\mathrm{d}h}{\mathrm{d}f}\right)_{\mathrm{{TH}}} \left( \frac{\mathrm{d}f}{\mathrm{d}t}\right)_\mathrm{TH}  
= \frac{6 k}{ \mathcal{Q}} \frac{\pi^4 m_2^2 f^3 R_1^5(f/2-\Omega_1)}{ G(m_1 + m_2)^2} .
\label{eq:AMev}
\end{equation}
In terms of the primary rotation frequency (spin) in Hz 
this is
\begin{equation}
 \left( \frac{\mathrm{d}\Omega_1}{\mathrm{d}t}  \right)_\mathrm{TH} = \frac{3 k}{ \mathcal{Q}} \frac{\pi^3 m_2^2 f^3 R_1^3(f/2-\Omega_1)}{ G m_1 r_g^2 (m_1 + m_2)^2} .
\label{eq:AMev20}
\end{equation}
In addition to accounting for angular momentum exchange, we must also account for the energy transferred into the primary, by the loss of orbital energy. The binary's orbital energy in terms of the gravitational wave frequency $f$ is given by
\begin{equation}
    E_\mathrm{bin} = -\frac{G m_1 m_2}{2a}= -\frac{ \pi^{2/3} G^{2/3} m_1 m_2}{2 \left( m_1 + m_2 \right)^{1/3}} f^{2/3},
    \label{eq:Ebin}
\end{equation}
while the rotational energy of the finite-sized primary is given by
\begin{equation}
    E_\mathrm{rot} = \frac{1}{2}I_1\Omega_1^2 \left(2 \pi \right)^2 = \frac{1}{2} m_1 r_g^2R_1^2 \Omega_1^2 \left(2 \pi \right)^2.
\end{equation}

The sum of these two $E_\mathrm{tot} = E_\mathrm{bin} + E_\mathrm{rot}$ determines the energy budget of the binary system. Given {Equation~(\ref{eq:dfdtTD0})}, the rate of change in the binary energy is
\begin{equation}
{ \left( \frac{\mathrm{d}{E}_\mathrm{bin}}{\mathrm{d}t} \right)_\mathrm{TH}} = \frac{\mathrm{d}E_\mathrm{bin}}{\mathrm{d}f} \left( \frac{\mathrm{d}f}{\mathrm{d}t} \right)_\mathrm{TH}= - \frac{6 k}{\mathcal{Q}} \frac{m_2^2 R_1^5 \pi^{5}}{ G (m_1+m_2)^{2}} {f^{4}\left(f/2-\Omega_1\right)}
\label{eq:Edbin}
\end{equation}
via Equation~(\ref{eq:Ebin}). Then for the change in rotational energy ${\dot{E}_\mathrm{rot}}$, the change is given by the increase in primary spin,
\begin{equation}
{ \left( \frac{\mathrm{d}{E}_\mathrm{rot}}{\mathrm{d}t} \right)_\mathrm{TH}} = \frac{\mathrm{d}E_\mathrm{rot}}{\mathrm{d}\Omega_1} \left( \frac{\mathrm{d}\Omega_1}{\mathrm{d}t} \right)_\mathrm{TH}=   \frac{3 k}{\mathcal{Q}} \frac{(2 \pi)^2 m_2^2 R_1^5 \pi^{3}}{ G (m_1+m_2)^{2}} {f^{3} \Omega_1\left(f/2-\Omega_1\right)}.
\label{eq:Edrot}
\end{equation}

Finally, the rate of change of total energy of the binary (binding energy and {primary} spin) due to decay by Equation~{(\ref{eq:dfdtTD0})} and subject to Equation~(\ref{eq:AMev}), is given by the sum of Equations~(\ref{eq:Edbin}) and (\ref{eq:Edrot})
\begin{equation}
{ \left( \frac{\mathrm{d}E_\mathrm{tot}}{\mathrm{d}t} \right)_\mathrm{TH}} = -12 \frac{k}{\mathcal{Q} }
 \frac{ \pi^5 R_1^5{m_2}^2 f^3}{G({m_1}+{m_2})^2}  \left( \left(f/2\right)^2- 2 \left(f/2\right) \Omega_1 + \Omega_1^2  \right),
 \label{eq:Etotdot0}
\end{equation}
which is equivalent to Equations A28--31 in \cite{Hut1981}.

\subsection{Primary response to tidal heating}
\label{sec:tempmodel}

Unlike angular momentum accounting in Equation~(\ref{eq:AMev}), energy in Equation~(\ref{eq:Etotdot}) is irreversibly lost (not exchanged) by dissipation (turbulent viscosity) in the primary. Specifically, orbital energy lost by Equation~(\ref{eq:dfdtTD}) heats the primary at a rate given by Equation~(\ref{eq:Etotdot}). This is accounted for {by an increase} in the surface luminosity of the {primary $L_1$,} which assuming a blackbody is given by
\begin{equation}
    L_1 = 4\pi\sigma R_1(T_1)^2 T_1^4
    \label{eq:lumeq0}
\end{equation}
where $\sigma = 5.67 \times 10^{-5}$ erg/s cm$^{-2}$ K$^{-4}$. A result of the energy flow from Equation~(\ref{eq:Etotdot0}) into the star {with $L_1$} is that the temperature $T_1$ and hence temperature dependent radius $R_1$ (Equation~(\ref{eq:RMTPanei})) increases. Th{is increase in the luminosity} is {given by}
\begin{equation}
\left( \frac{\mathrm{d}L_1}{\mathrm{d}t}\right)=  -\left( \frac{\mathrm{d}\dot{E}_\mathrm{tot}}{\mathrm{d}t}\right)_\mathrm{{TH}}= - \left( \frac{\mathrm{d}\dot{E}_\mathrm{tot}}{\mathrm{d}f}\right) \left( \frac{\mathrm{d}f}{\mathrm{d}t}\right)_\mathrm{TH}
\end{equation}
via {Equations~(\ref{eq:dfdtTD0})} and {(\ref{eq:Etotdot0})}. Then following {Equation~(\ref{eq:lumeq0})}, the temperature of the primary changes with {$L_1$} according to
\begin{equation}
    \left( \frac{\mathrm{d}T_1}{\mathrm{d}{{L}_1}}\right)  =\left( {8 \pi  \sigma  T_1^3 R(T_1) \left(T_1 R'(T_1)+2 R(T_1)\right)} \right)^{-1},
\end{equation}
noting the additional temperature dependence through the radius. Finally we write the time evolution of the temperature:
\begin{equation}
\frac{\mathrm{d}T_1}{\mathrm{d}t} = \frac{\mathrm{d}L_1}{\mathrm{d}t} \frac{\mathrm{d}T_1}{\mathrm{d}L_1}
 = 
\frac{135 \pi^{25/3} R_1^9{m_2}^3 f^{19/3}}{G^{8/3}m_1({m_1}+{m_2})^{11/3}} \left(\frac{k}{\mathcal{Q} } \right)^2   \left(\frac{f}{2} -\frac{3}{5}\Omega_1 \right)  \left(\frac{f}{2} - \Omega_1\right)^2
 \times \left(\sigma T^3 \left(2 R_1 + \frac{\mathrm{d}R_1}{\mathrm{d}T_1} T_1 \right)\right)^{-1}.
 \label{eq:dT1dt0}
\end{equation}

\bibliographystyle{aasjournal}

\bibliography{ELMtides}

\end{document}